\PassOptionsToPackage{colorlinks=true,linkcolor=black,anchorcolor=black,citecolor=black,filecolor=black,menucolor=black,runcolor=black,urlcolor=black}{hyperref}

\documentclass[letterpaper, 10 pt, journal]{IEEEtran}
\IEEEoverridecommandlockouts

\usepackage[utf8]{inputenc}
\usepackage{textcomp}
\usepackage{multirow}
\usepackage{graphicx}
\usepackage{mathtools}
\usepackage{amssymb}
\usepackage{amsmath}
\usepackage{bm}
\usepackage{subcaption}
\usepackage{siunitx}
\usepackage{longtable,tabularx}
\usepackage{float}
\usepackage{booktabs}
\usepackage{bookmark}
\usepackage{etoolbox}
\usepackage{multirow}
\usepackage{subcaption}
\usepackage{xcolor}

\usepackage{url}
\usepackage{hyperref}
\usepackage{BernsteinStyle}

\usepackage{url}
\usepackage{tikz}
\usepackage{pgfplots}
\usepackage{accents}
\usepackage{comment}

\usetikzlibrary{calc}
\usetikzlibrary{shapes, arrows}
\usetikzlibrary{positioning}
\usepackage{color}
\usepackage{circuitikz}
\usepackage{etoolbox}

\usetikzlibrary{shapes,arrows,calc,positioning}
\tikzstyle{bigblock} = [draw, fill=blue!20, rectangle, 
    minimum height=6em, minimum width=8em]
\tikzstyle{medblock} = [draw, fill=blue!20, rectangle, 
    minimum height=4em, minimum width=4em]    
\tikzstyle{mux} = [draw, fill=black!20, rectangle, 
    minimum height=5em, minimum width=0.1em]    
\tikzstyle{smallblock} = [draw, fill=blue!20, rectangle, 
    minimum height=2.5em, minimum width=4em]
\tikzstyle{sum} = [draw, fill=blue!20, circle, node distance=1cm]
\tikzstyle{signal} = [coordinate]
\tikzstyle{pinstyle} = [pin edge={to-,thin,black}]
\tikzstyle{block} = [draw, fill=blue!20, rectangle, 
    minimum height=3em, minimum width=6em]
\tikzstyle{blockS} = [draw, fill=blue!20, rectangle, 
    minimum height=3em, minimum width=4em]  
\tikzstyle{sum} = [draw, fill=blue!20, circle, node distance=1.5cm]
\tikzstyle{gain} = [draw, fill=blue!20, regular polygon, regular polygon sides = 3, node distance=1.25cm, shape border rotate = -90]
\tikzstyle{mult} = [draw, fill=blue!20, circle, inner sep=0pt, minimum size=0.2cm,]
\tikzstyle{input} = [coordinate]
\tikzstyle{output} = [coordinate]

\pgfplotsset{compat=1.15}

\title{\LARGE Self-Excited Dynamics of Discrete-Time Lur'e Models \\
with Affinely Constrained, Piecewise-C$^1$ Feedback Nonlinearities}

\author{\large Juan Paredes, Omran Kouba, and Dennis S. Bernstein
\thanks{Juan Paredes, and Dennis S. Bernstein are with the Department of Aerospace Engineering, University of Michigan, Ann Arbor, MI, USA. {\tt\small \{jparedes, dsbaero\}@umich.edu}  Omran Kouba is with the Department of Mathematics in the Higher Institute of Applied Sciences and Technology, Damascus, Syria. {\tt\small omran\_kouba@hiast.edu.sy}}
}

\begin{document}

\maketitle

\begin{abstract}
Self-excited systems (SES) arise in numerous applications, such as fluid-structure interaction, combustion, and biochemical systems. 
In support of system identification and digital control of SES, this paper analyzes discrete-time Lur'e models with affinely constrained, piecewise-C$^1$ feedback nonlinearities.
The main result provides sufficient conditions under which a discrete-time  Lur'e model is self-excited in the  sense that its response is 1) bounded for all initial conditions, and 2) nonconvergent for almost all initial conditions.
\end{abstract}

\section{Introduction}

{\it Self-excited systems} (SES) have the property that constant inputs lead to oscillatory outputs \cite{Ding2010,JENKINS2013167}.
The diversity of applications in which SES arise is vast, and encompasses
fluid-structure interaction \cite{coller,cesnik},
thermoacoustic oscillations \cite{awad_1986,epperlein2015,chen_2016},
and chemical and biochemical systems \cite{goldbeter,chance,gray_scott_1990}.
Not surprisingly, extensive effort has been devoted to modeling and controlling SES \cite{livne2018,dowling2005,yi2007}.
SES are also used for controller tuning;  
for PID control, a relay inserted inside a servo loop  induces limit-cycle oscillations, which are used to identify the crossover frequency \cite{hang2002}.

Control of SES requires analytical and empirical models;
the present paper is motivated by the latter need.
System identification for SES based on continuous-time Lur'e models is considered in \cite{Savaresi}.
Alternatively, for sampled-data control, system identification for SES based on discrete-time Lur'e models is considered in \cite{JuanIJC2022}. 
In support of discrete-time system identification and sampled-data control of SES, the present paper focuses on discrete-time Lur'e models of SES.

A Lur'e model consists of linear dynamics with memoryless nonlinear feedback.
The stability of Lur'e models is a classical problem, expressed by the Aizerman conjecture for sector-bounded nonlinearities 
\cite{aizerman1949, bragin2011, leonov2013}.
Although the Aizerman conjecture is false, the stability of Lur'e models has been widely studied in both continuous time  
\cite{kalman1963,aizerman1964,yakubovich1973freq,liu2010,sarkans2015, yakubovich1973,Tomberg1989, chua1990, arcak2002, arcak2002Lurie, sepulchre2005,efimov2009}
and discrete time 
\cite{haddad1993,park1998,ahmad2012,gonzaga2012,ahmad2014,park2015,sarkans2016,bertolin2021,bertolin2022,guiver2022,drummond2023,vrasvan98,amico2001,seiler2020}.

In contrast to stable behavior, many SES are modeled by Lur'e models that have unstable equilibria and bounded response. 
A classical example is the Lur'e model of a Rijke tube, in which acoustic waves interact through feedback with the flame dynamics to produce thermoacoustic oscillations \cite{epperlein2015,chen_2016}.
Self-excited oscillations in continuous-time Lur'e models
have been studied in \cite{yakubovich1973,Tomberg1989,arcak2002,sepulchre2005,efimov2009}.
In particular, using the bounded real lemma, continuous-time Lur'e models with superlinear feedback and minimum-phase linear dynamics with relative degree 1 or 2 are shown in \cite{arcak2002} to possess bounded solutions.
Related results are given in \cite{sepulchre2005} based on dissipativity theory as well as in \cite{efimov2009} using Lyapunov methods.

In contrast to \cite{yakubovich1973,Tomberg1989,arcak2002,sepulchre2005,efimov2009}, the present paper focuses on discrete-time, self-excited Lur'e models, with the property that, for all constant inputs, the response is 1) bounded for all initial conditions, and 2) nonconvergent for almost all initial conditions.
The main contribution of the present paper is sufficient conditions for this behavior for a specific class of nonlinear feedback functions.
The analogous property for continuous-time Lur'e models is not addressed in the literature.

It is important to stress the distinctions between continuous-time and discrete-time Lur'e models that exhibit self-excited behavior.
In particular, since superlinear feedback has unbounded gain, the linear dynamics of a continuous-time Lur'e model must be high-gain stable.
From a root locus perspective, this means that the linear dynamics must be minimum phase, the relative degree cannot exceed 2, and, when the relative degree is 2, the root locus center must lie in the open left half plane.
These conditions, which are invoked in \cite{arcak2002} for continuous-time dynamics, do not imply high-gain stability for discrete-time systems with strictly proper linear dynamics.
As discussed in \cite{sarkans2016}, bounded response of a discrete-time Lur'e model with superlinear feedback requires positive-real, and thus relative-degree-zero, linear dynamics.
Superlinear feedback is thus incompatible with discrete-time Lur'e models of SES.

Table \ref{Lure_ref_tab} categorizes some of the literature on continuous-time (CT) and discrete-time (DT) Lur'e models in terms of asymptotically stable response and bounded, nonconvergent response.
The most relevant among these works to the present paper are \cite{vrasvan98,amico2001,seiler2020} on discrete-time Lur'e models that have bounded, nonconvergent response.
In particular, 
\begin{itemize}
    \item \cite{vrasvan98} extends the results of \cite{yakubovich1973} to discrete-time systems, and considers a discrete-time Lur'e model with a sector-bounded nonlinearity that induces oscillations.
    \item \cite{amico2001} provides a graphical tool based on Hopf bifurcation for analyzing discrete-time Lur'e models with a smooth  nonlinearity that yields a self-excited response.
    \item \cite{seiler2020} considers a discrete-time Lur'e model and provides sufficient conditions for the existence of a slope-restricted nonlinearity that yields a self-excited response.
    Under these conditions, the set of initial conditions that give rise to the self-excited response may have measure zero.
\end{itemize}
None of these works, however, provides conditions under which a discrete-time Lur'e model is self-excited in the sense of the present paper.

\begin{table}[h]
\caption{Lur'e Model Literature}
\label{Lure_ref_tab}
\centering
\renewcommand{\arraystretch}{1.2}
\resizebox{\columnwidth}{!}{%
\begin{tabular}{c|c|c|}
\cline{2-3}
 & $\begin{array}{c} \mbox{\textbf{Asymptotically }} \\ \mbox{ \textbf{Stable}}           \end{array}$ & $\begin{array}{c}\mbox{\textbf{Bounded and}} \\ \mbox{\textbf{Nonconvergent}}\end{array}$\\
\hline
\multicolumn{1}{|c|}{\textbf{CT}}  &  \cite{kalman1963,aizerman1964,yakubovich1973freq,liu2010,sarkans2015}
& $\begin{array}{c} \cite{yakubovich1973,Tomberg1989,chua1990,arcak2002} \\ \cite{arcak2002Lurie, sepulchre2005,efimov2009} \end{array}$ \\
\hline
\multicolumn{1}{|c|}{\textbf{DT}}  & $ \begin{array}{c} \cite{haddad1993,park1998,ahmad2012,gonzaga2012,ahmad2014,park2015} \\ \cite{sarkans2016,bertolin2021,bertolin2022,guiver2022,drummond2023} \end{array}$
& \cite{vrasvan98,amico2001,seiler2020} \\
\hline
\end{tabular}
}
\end{table}

In order to address the special features of self-excited discrete-time Lur'e models, the main contribution of the present paper is to prove that a class of discrete-time Lur'e models with {\it affinely constrained} feedback are self-excited in the  sense that 1) all trajectories are bounded and 2) the set of initial conditions for which the state trajectory is convergent has measure zero.
Although an affinely constrained function need not be bounded or even sector-bounded, it must have linear growth, thus ruling out superlinear nonlinearities, as necessitated by the fact that discrete-time strictly proper linear systems are not high-gain stable.
By bounding the feedback gain, the linear-growth assumption enables self-oscillating discrete-time Lur'e models with unbounded feedback nonlinearities.
As a benefit of this setting, the linear discrete-time dynamics of the Lur'e model need not be minimum phase, which is assumed in \cite{arcak2002} for continuous-time systems.

An additional novel feature of the discrete-time Lur'e model considered in this paper is the structural assumption that the linear dynamics possess a zero at 1.
This assumption, which places a washout filter in the loop, blocks the DC component arising from the constant exogenous input to the system and ensures that the nonlinear closed-loop system have a unique equilibrium for each constant, exogenous input.
Most importantly, this property prevents the Lur'e model from having an additional equilibrium with a nontrivial domain of attraction.

The main contribution of the paper is Theorem \ref{TheoConvergence1}, which provides conditions under which the set of initial conditions for which the trajectories of the Lur'e model are convergent has measure zero.
This result is applicable to discrete-time Lur'e models with piecewise-C$^1$ nonlinearities for which the Jacobian of the closed-loop dynamics may be singular on a set of measure zero.
The need to consider piecewise-C$^1$ nonlinearities is motivated by their role in nonlinear system identification \cite{JuanIJC2022, vanpelt2001, dolanc2005, voros2007}.
Under the stronger assumptions of C$^1$ nonlinearities and everywhere-nonsingular Jacobian, Theorem 2 in \cite{lee2019} is applicable.
Theorem \ref{TheoConvergence1} thus extends Theorem 2 in \cite{lee2019} to the case where the nonlinearity is  piecewise-C$^1$ (and thus not necessarily C$^1$) and the Jacobian of the closed-loop dynamics may be singular on a set of measure zero.
Finally, Theorem \ref{TheoConvergence} has no counterpart in \cite{DTLACC}, and thus the results in the present paper provide a substantial extension of \cite{DTLACC}.

\begin{figure}[H]
	\centering
	\resizebox{0.7\columnwidth}{!}{%
		\begin{tikzpicture}[>={stealth'}, line width = 0.25mm]
		\node[draw = none] at (0,0) (prop23) {Proposition \ref{propequilibrium}};
		\node[draw = none, below = 2 em of prop23.center] (prop26) {Proposition \ref{PropConvergence}};
		\node[draw = none, above right = 0.7em and 7em of prop23.center] (lem32) {Lemma \ref{lem:un_meas0_closed}};
		\node[draw = none, below = 2 em of lem32.center] (prop34) {Proposition \ref{prop:DEmeasMu0}};
		\node[draw = none, below = 2 em of prop34.center] (lem35) {Lemma \ref{lem:finvMeas0}};
		\node[draw = none, below right =3.5em and 2.25em of prop26.center] (theo36) {Theorem \ref{TheoConvergence1}};
		\node[draw = none, below = 2em of theo36.center] (theo39) {Theorem \ref{TheoConvergence}};
		\draw[->] (prop23)--(prop26);
		\draw[->] (lem32)--(prop34);
		\draw[->] (prop34)--(lem35);
		\draw[->] (prop26)--(theo36);
		\draw[->] (lem35)--(theo36);
		\draw[->] (theo36)--(theo39);
		\draw[->] (prop23.west) -| ([xshift = -7em]theo36.west) -- (theo36.west);
		\end{tikzpicture}
	}
	\caption{Result dependencies. }
	\label{statement_flowchart}
\end{figure}
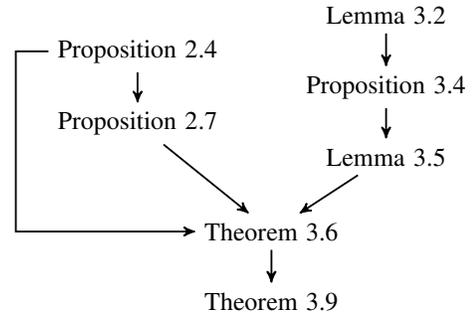

The contents of the paper are as follows.
Section \ref{sec:LureAnalysis} introduces the discrete-time Lur'e model, which involves asymptotically stable linear dynamics in feedback with a memoryless nonlinearity, and analyzes its equilibrium properties.
Section \ref{sec:LureConvergence} defines affinely constrained nonlinearities and provides sufficient conditions under which the discrete-time Lur'e model possesses a bounded, nonconvergent response for almost all initial conditions.
In particular, Theorem \ref{TheoConvergence} provides a sufficient condition for the Lur'e model to be self-excited.
Theorem \ref{TheoConvergence} depends on Theorem \ref{TheoConvergence1}, which provides conditions under which the set of initial conditions for which the state trajectory converges has measure zero.
In the case where the feedback nonlinearity is C$^1$ and the Jacobian of the closed-loop dynamics is nonsingular at all points, Theorem \ref{TheoConvergence1} follows from Theorem 2 in \cite{lee2019}.
The case where the feedback nonlinearity is only piecewise C$^1$ is required for system identification as considered in \cite{JuanIJC2022}, where the identified feedback nonlinearity is constructed to be piecewise affine. 
Finally, Section \ref{sec:LureNumericalExamples} presents numerical examples that illustrate the conditions for self-excitation presented in Section \ref{sec:LureConvergence}.
Figure \ref{statement_flowchart} shows the dependencies of the results in this paper.

{\bf Nomenclature and terminology.}  $\BBR \isdef (-\infty,$ $\infty),$ $\BBN \isdef \{1, 2, \ldots\},$ $\BBN_0 \isdef \{0, 1, 2, \ldots\},$ $\BBC$ denotes the complex numbers,
$\Vert\cdot\Vert$ denotes the Euclidean norm on $\BBC^n,$ and $\bfz\in\BBC$ denotes the Z-transform variable.
For $\SG \subseteq \BBR^n,$ ${\rm acc} (\SG)$ denotes the set of accumulation points of $\SG$ (Definition \ref{def:acc_point}).
For $\SG \subseteq \BBR^n,$ $\dim(\SG)$ denotes the dimension of $\SG,$
and, for (Lebesgue) measurable $\SG \subseteq \BBR^n,$ $\mu(\SG)$ denotes the measure of  $\SG.$
For $x \in \BBR^n$ and  $\varepsilon > 0,$ $\BBB_\varepsilon (x)$ denotes the open ball of radius $\varepsilon$ centered at $x$.
Positive-definite matrices are assumed to be symmetric.
For $A\in\BBR^{n\times n},$ $\spr(A)$ denotes the spectral radius of $A,$ 
$\Vert A \Vert$ denotes the maximum singular value of $A,$
and, if $A$ is positive definite, then $\lambda_{\rm min}(A)$ denotes the eigenvalue of $A$ of minimum magnitude and $\lambda_{\rm max}(A)$ denotes the eigenvalue of $A$ of maximum magnitude.
The terminology ``$\lim_{k\to\infty}\alpha_k$ exists'' implies that the indicated limit is finite.

\section{Analysis of the Lur'e Model}
\label{sec:LureAnalysis}

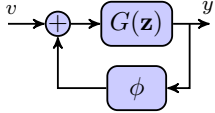
\begin{figure}[H]
	\centering
	\resizebox{0.35\columnwidth}{!}{%
		\begin{tikzpicture}[>={stealth'}, line width = 0.25mm]
		\node [input, name=input] {};
		\node [sum, right = 0.5cm of input] (sum1) {};
		\node[draw = white] at (sum1.center) {$+$};
		\node [smallblock,  rounded corners, right = 0.4cm of sum1, minimum height = 0.6cm, minimum width = 0.8cm] (system) {$  G(\bfz)$};
		\node [smallblock, rounded corners, below = 0.25cm of system, minimum height = 0.6cm, minimum width = 0.8cm] (satF) {$\phi$};
		\draw [->] (input) -- node[name=usys, xshift = -0.2cm, yshift = 0.2cm] {\footnotesize$v$} (sum1.west);
		\draw[->] (sum1.east) -- (system.west);
		\node [output, right = 0.5cm of system] (output) {};
		\draw [->] (system) -- node [name=y,near end, xshift = -0.15cm]{} node [near end, above, xshift = 0.1cm] {\footnotesize$y$}(output);
		\draw [->] (y.center) |- (satF.east);
		\draw [->] (satF.west) -| (sum1.south);
		\end{tikzpicture}
	}
	\caption{Discrete-time Lur'e model.}
	\label{DT_Lure_blk_diag}
\end{figure}

Let $G(\bfz) = C(\bfz I - A)^{-1}B$ be a strictly proper, discrete-time SISO transfer function with $n$th-order minimal realization $(A, B, C)$ and state $x_k \in \BBR^n$ at step $k,$ let $\phi\colon \BBR \to \BBR$, and let $v\in\BBR.$
Then, for all $k\ge0,$ the discrete-time Lur'e model in Fig. \ref{DT_Lure_blk_diag} has the closed-loop dynamics
\begin{align}
    x_{k+1} &= A x_k +  B (\phi(y_k) + v), \label{xLure}\\
    y_k &= C x_k,\label{yLure}
\end{align}
and thus 
\begin{equation}
    y_k = CA^k x_0 +  \sum_{i=0}^{k-1} C A^{k-1-i} B (\phi(y_i) + v).\label{ysoln}
\end{equation}
Note that \eqref{xLure}, \eqref{yLure} can be written as
\begin{align}
    x_{k+1} = f(x_k), \label{xisfatx}
\end{align}
where $f(x) \isdef A x +  B (\phi(Cx) + v).$
Henceforth, we assume that $n\ge 2.$

\begin{defin} \label{lureDefin}
    \eqref{xLure}, \eqref{yLure} is {\it self-excited} if, for all $v\in\BBR,$ the following statements hold:
    \begin{enumerate}
        \item For all $x_0\in\BBR^n,$ $(x_k)_{k=1}^\infty$ is bounded.
        \item  For almost all $x_0\in\BBR^n,$ $\lim_{k\to\infty}x_k$ does not exist. 
    \end{enumerate}
\end{defin}

Note that {\it ii}) holds if and only if $\{x_0\in\BBR^n \colon \lim_{k\to\infty}x_k \mbox{ exists}\}$ has measure zero.
The following result concerns the measure of the set of initial conditions for which the output converges.

\begin{prop}
Assume that $\spr(A) < 1,$ and, for all $v\in\BBR,$  $\{x_0\in\BBR^n \colon \lim_{k\to\infty}x_k \mbox{ exists}\}$ has measure zero. Then, $\{x_0\in\BBR^n \colon \lim_{k\to\infty}y_k \mbox{ exists}\}$ has measure zero.
\end{prop}
{\it Proof:}
Suppose that $X_0 \isdef \{x_0\in\BBR^n \colon \lim_{k\to\infty}y_k \mbox{ exists}\}$ has positive measure. 
For all $x_0 \in X_0,$ $\lim_{k\to\infty} (\phi(y_k) + v)$ exists, and thus, since $\spr(A) < 1,$ it follows from \eqref{xLure} and input-to-state stability for linear time-invariant discrete-time systems
\cite[Example 3.4]{ISSCT}
that, for all $x_0 \in X_0,$ $\lim_{k\to\infty} x_k$ exists, which is a contradiction.
\hfill $\square$

\begin{defin}
    $x \in \BBR^n$ is an {\it equilibrium} of \eqref{xLure}, \eqref{yLure} if $x$ is a fixed point of $f,$ that is,
    \begin{equation}
        x = A x +  B (\phi(Cx)+v).\label{xeeqn}
    \end{equation}
\end{defin}

When $I-A$ is nonsingular, define
\begin{equation} \label{x_rme_def}
    x_\rme \isdef (I - A)^{-1} B v
\end{equation}
and note that
\begin{equation}
Cx_\rme = G(1)v.\label{CxeisG1v}
\end{equation}
The following result establishes useful properties of $G$ and $\phi$.

\begin{prop} \label{propequilibrium}
Assume that $I-A$ is nonsingular.  Then, the following statements hold:
\begin{enumerate}
    \item $x \in \BBR^n$ is an equilibrium of \eqref{xLure}, \eqref{yLure} if and only if
    \begin{equation}
        x =  (I-A)^{-1}B (\phi(C x) + v).\label{xeG0}
    \end{equation}
    \item If $x \in \BBR^n$ is an equilibrium of \eqref{xLure}, \eqref{yLure}, then the following statements hold:\\
    $a)$  $C x =  G(1) (\phi(C x) + v).$\\
    $b)$  $\phi(Cx) = -v$ if and only if $x=0$.\\  
    $c)$  If $G(1) = 0$, then $Cx=0$ and $x = (I-A)^{-1}B (\phi(0) + v)$ is the unique equilibrium of \eqref{xLure}, \eqref{yLure}.\\
    $d)$  If $Cx=0,$ then either $G(1)=0$ or $v = -\phi(0)$. \\
    $e)$  If $\phi(Cx)=0$, then $x = x_\rme$.   
    \item   The following statements are equivalent:\\
    $a)$  $x_\rme$ is an equilibrium of \eqref{xLure}, \eqref{yLure}.\\
    $b)$  $\phi(Cx_\rme)=0$.\\
    $c)$  $\phi(G(1)v)=0$.
    \item Assume that $G(1) \neq 0.$ Then, the following statements are equivalent:\\
    $a)$  $x_\rme$ is an equilibrium of \eqref{xLure}, \eqref{yLure}.\\
    $b)$  $\phi(Cx_\rme)=0$.\\
    $c)$  $v\in\frac{1}{G(1)}\phi^{-1}(\{0\}).$
    \item  Assume that $G(1) = 0.$  Then, the following statements are equivalent:\\
    $a)$  $\phi(0)=0.$\\
    $b)$  $x_\rme$ is an equilibrium of \eqref{xLure}, \eqref{yLure}.\\
    $c)$  $x_\rme$ is the unique equilibrium of \eqref{xLure}, \eqref{yLure}.
\end{enumerate}
\end{prop}

The proof of Proposition \ref{propequilibrium} is given in the appendix.
Note that the converse of  Proposition \ref{propequilibrium}$ii)e)$ is true and is given by $iii)$.

\begin{exam}\label{ex_conv_1}
{\it 
This example shows that the converse of Proposition \ref{propequilibrium}$ii)d)$ is false.
}
Let $v = 0,$ $\phi(y) = \tanh(y),$ and $G (\bfz) = 1/(\bfz^2 - \bfz + 0.5)$, with minimal realization
\begin{equation*}
    A = \begin{bmatrix} 1 & -0.5 \\ 1 & 0 \end{bmatrix}, \ B = \begin{bmatrix} 1 \\ 0 \end{bmatrix}, \ C = \begin{bmatrix} 0 & 1 \end{bmatrix}.
\end{equation*}
Note that $\phi(0)=0$, $G(1)\ne0,$ and the Lur'e model \eqref{xLure}, \eqref{yLure} has three equilibria, namely, $x_{\rme, 1} = [0 \ \ 0]^\rmT,$ $x_{\rme, 2} = [1.91501 \ \ 1.91501]^\rmT,$ and $x_{\rme, 3} = [-1.91501 \ \ -\mspace{-3mu}1.91501]^\rmT.$ 
Hence, $\phi(C x_{\rme,1}) = 0 = -v,$ $\phi(C x_{\rme,2}) \ne 0,$ and $\phi(C x_{\rme,3}) \ne 0.$
Therefore, consistent with Proposition \ref{propequilibrium}$ii)b)$, $ x_{\rme,1}=0,$ and 
$x_{\rme,2}$ and  $x_{\rme,3}$ are both nonzero.
Furthermore,  $G(1) \ne 0$, $v = -\phi(0) = 0,$ and $C x_{\rme,2}$ and $C x_{\rme,3}$ are nonzero.
Since Proposition \ref{propequilibrium}$ii)c)$ implies that, for all equilibria $x\in\BBR^n$ of \eqref{xLure}, \eqref{yLure},
$G(1) = 0$ implies $Cx=0,$ it follows that the converse of  Proposition \ref{propequilibrium}$ii)d)$ is false.
\hfill$\huge\Diamond$
\end{exam}

\begin{exam}\label{ex_conv_2}
{\it
This example shows that the converse of Proposition \ref{propequilibrium}$v)$ is false.
}
Let $v= 0,$ $\phi(y) = y,$ and $G$ be as in Example \ref{ex_conv_1}, and let $x$ be an equilibrium of \eqref{xLure}, \eqref{yLure}. 
It follows from \eqref{xeeqn} that
    $x = (A + BC) x.$
Since $I - A - BC$ is nonsingular, it follows that $x = x_\rme = 0$ is the unique equilibrium of \eqref{xLure}, \eqref{yLure}.
Since $G(1)$ is nonzero and $\phi(0) = 0$, it follows that the converse of Proposition \ref{propequilibrium}$ii)c)$ is false.
Furthermore, although $a),$ $b),$ and $c)$ of Proposition \ref{propequilibrium}$v)$ are satisfied, $G(1)$ is nonzero.
\hfill$\huge\Diamond$
\end{exam}

In the following result, the first statement implies that every convergent state trajectory of \eqref{xLure}, \eqref{yLure} converges to an equilibrium solution.
Under stronger conditions, the second statement implies that every convergent state trajectory of \eqref{xLure}, \eqref{yLure} converges to the unique equilibrium solution given by \eqref{x_rme_def}.

\begin{prop} \label{PropConvergence}
Assume that $I-A$ is nonsingular and $\phi$ is continuous.
Then, the following statements hold:
\begin{enumerate}
    \item If $x_\infty\isdef \lim_{k\to\infty}x_k$ exists, then $x_\infty$ is an equilibrium of \eqref{xLure}, \eqref{yLure}.
    \item Assume that $G(1)=0$ and $\phi(0) = 0.$ 
    Then, the following statements hold:\\
    $a)$ If $x_\infty\isdef \lim_{k\to\infty}x_k$ exists, then $x_\infty = x_\rme.$ \\
    $b)$ $\{x_0\colon \lim_{k\to\infty}x_k \mbox{  exists}\}=\{x_0\colon \lim_{k\to\infty}x_k=x_\rme\}.$
\end{enumerate}
\end{prop}

{\it Proof:}
To prove $i)$, note that, since $\phi$ is continuous, it follows that $f$ is continuous.
Hence, \eqref{xisfatx} implies that $x_\infty = \lim_{k \to \infty}x_k = \lim_{k \to \infty} f(x_k) = f(x_\infty).$

To prove $ii)a)$, note that $i)$ implies that $x_\infty$ is an equilibrium of \eqref{xLure}, \eqref{yLure}.
Since $G(1)=0$ and $\phi(0) = 0,$ Proposition \ref{propequilibrium}$v)$ implies that $x_\rme$ is the unique equilibrium of \eqref{xLure}, \eqref{yLure}.
Hence, $x_\infty = x_\rme.$

To prove $ii)b),$ note that ``$\subseteq$''  follows from $ii)a)$.
Finally, ``$\supseteq$'' is immediate.
\hfill $\square$

\section{Self-Excited Dynamics of the Lur'e Model} \label{sec:LureConvergence}

This section presents sufficient conditions under which the Lur'e model \eqref{xLure}, \eqref{yLure} with an affinely constrained nonlinearity is self-excited.

\subsection{Preliminary Results}

\begin{defin}\label{def:acc_point}
Let $\SB\subseteq\BBR^n.$  Then, $z\in\SB$ is an {\it isolated point of $\SB$} if there exists $\varepsilon>0$ such that $\BBB_\varepsilon(z) \cap(\SB\backslash\{z\}) = \varnothing.$
Furthermore, $z\in\BBR^n$ is an {\it accumulation point} of $\SB$ if, for all $\varepsilon > 0,$
$\BBB_\varepsilon(z)\cap(\SB\backslash\{z\})\ne\varnothing.$
The set of accumulation points of $\SB$ is denoted by ${\rm acc}(\SB)$, and the set of isolated points of $\SB$ is denoted by ${\rm iso}(\SB)$.
\end{defin}

It can be seen that $z \in {\rm acc}(\SB)$
if and only if there exists $(x_i)_{i=1}^\infty\subseteq \SB\backslash \{z\}$ such that $\lim_{i\to\infty}x_i = z.$
Note that $z\in{\rm acc}(\SB)$ need not be an element of $\SB.$
In fact, $\cl(\SB)\backslash\SB \subseteq \cl(\SB)\backslash{\rm iso}(\SB) = {\rm acc}(\SB),$
and thus ${\rm acc}(\SB)=\varnothing$ if and only if $\SB = {\rm iso}(\SB).$

\begin{lem}\label{lem:un_meas0_closed}
    Let $\SA \subseteq \BBR$, assume that ${\rm acc}(\SA) = \varnothing,$ and define $\SB \isdef \{x \in \BBR^n \colon Cx \in\SA\}$. 
    Then, the following statements hold:
    \begin{enumerate}
        \item $\SB$ has measure zero.
        \item $\SB$ is closed. 
    \end{enumerate}
\end{lem}

{\it Proof:}
Both statements are true when $\SA$ is empty; hence assume that $\SA$ is not empty.
To prove $i$), note that $\SB$ is the union of hyperplanes, each of which has measure zero.
Since ${\rm acc}(\SA) = \varnothing,$ 
$\SA$ is countable, and thus $\SB$ is a countable union of sets, each with measure zero.
Therefore, $\SB$ has measure zero. 
To prove $ii$), note that, since ${\rm acc}(\SA)=\varnothing,$ it follows that
$\SA = {\rm iso}(\SA),$ and thus $\SA$ is closed. 
Hence, $\SB$ is closed. \hfill $\square$

\subsection{Piecewise-C$^1$ Functions}

\begin{defin}
$\phi$ is {\it piecewise continuously differentiable} (PWC$^1$) if the following conditions hold:
\begin{enumerate}
    \item $\phi$ is continuous.
    \item Define $\SR \isdef \{y \in \BBR \colon \phi'(y)$ exists and $\phi'$ is continuous at $y\}.$ 
    Then, $\SSS \isdef \BBR\backslash\SR$ has no accumulation points.%
    \item For all $y \in \SSS,$ $\lim_{t \uparrow 0} \phi'(y + t)$ and $\lim_{t \downarrow 0} \phi'(y + t)$ exist.
\end{enumerate}
\end{defin}

Note that, if $\phi$ is C$^1,$ then $\SSS = \varnothing.$

As an example, consider $\phi(y) = y^2\sin(1/y)$ for $y\ne0$ and $\phi(0) = 0.$
Then, $\phi'(y) = 2y\sin(1/y)-\cos(1/y)$ for $y\ne0$ and $\phi'(0) = 0.$
Hence, $\SR = \BBR\backslash\{0\}$ and $\SSS = \{0\}.$
However, neither $\lim_{t \uparrow 0} \phi'(t)$ nor $\lim_{t \downarrow 0} \phi'(t)$ exists, and thus $\phi$ is not PWC$^1$.

It can be shown that, if $\phi'(y)$, $\lim_{t \uparrow 0} \phi'(y + t),$ and $\lim_{t \downarrow 0} \phi'(y + t)$ exist, then $\phi'(y) = \lim_{t \uparrow 0} \phi'(y + t) = \lim_{t \downarrow 0} \phi'(y + t)$, and thus $\phi'$ is continuous at $y.$  
Therefore, if $\phi$ is PWC$^1$ and $y\in\SSS,$ then $\phi'(y)$ does not exist.
Furthermore, {\it ii}) holds if and only if each bounded subset of $\BBR$ contains a finite number of elements of $\SSS.$

Assume that $\phi$ is PWC$^1$.
Then, define $\SD \isdef \{x \in \BBR^n \colon C x \in \SR\}$ and
$\SE \isdef \{x \in \BBR^n \colon C x \in \SSS\} = \BBR^n \backslash \SD$
so that $\SR=C\SD$ and $\SSS=C\SE.$
If $x\in\SD,$ then $f'(x) = A + \phi'(Cx) BC.$ 
Note that, in the case where $G(1) = 0,$ $f'(x_\rme) = f'(0) = A+\phi'(0)BC.$
Finally, define 
\begin{equation}
    \SD_0 \isdef \{ x \in \SD \colon f'(x) \ \mbox{ is singular} \}
\end{equation}
and
\begin{equation}
    \SR_0 \isdef C\SD_0.
\end{equation}
It thus follows that 
\begin{equation}
    \SR_0 = \{y\in\SR \colon A + \phi'(y) BC \mbox{ is singular} \}\subseteq\SR.
\end{equation}

\begin{prop}\label{prop:DEmeasMu0}
    Assume that $\phi$ is PWC$^1$
    and ${\rm acc}(\SR_0) = \varnothing.$
    Then, $\SD_0$ and $\SE$ are closed and have measure zero. 
\end{prop}

{\it Proof:}
Write
\begin{align*}
    \SD_0 &= \bigcup_{y\in \SR_0} \{x \in \BBR^n \colon Cx = y\},\\
    \SE &= \bigcup_{y \in \SSS} \{x \in \BBR^n \colon Cx = y\}.
\end{align*}
Since ${\rm acc}(\SR_0) = {\rm acc}(\SSS) = \varnothing,$
$i$) and $ii$) of Lemma \ref{lem:un_meas0_closed} imply that $\SD_0$ and $\SE$ are closed and have measure zero.
\hfill $\square$

Next, define
$f^1\isdef f$ and, for all $k\ge 1,$ $f^{k+1} \isdef f\circ f^k.$
Furthermore, for all $\SM\subseteq\BBR^n,$ define $f^{-1} (\SM) \isdef \{x \in \BBR^n \colon f(x) \in\SM\}$ and, for all $k\ge1,$   $f^{-k-1} (\SM) \isdef f^{-1}(f^{-k}(\SM)).$

\begin{lem} \label{lem:finvMeas0}
    Assume that $\phi$ is PWC$^1$ and
    ${\rm acc}(\SR_0) = \varnothing,$
    and let $\SM \subset \BBR^n$ have measure zero.  
    Then, for all $k\ge1,$ $\mu(f^{-k}(\SM))=0.$
\end{lem}

{\it Proof:}
    Proposition \ref{prop:DEmeasMu0} implies that $\SD_0$ and $\SE$ are closed, and thus $\SU \isdef \BBR^n\backslash(\SD_0 \cup \SE)$ is open.
    Next, since $\SU \cap (\SD_0 \cup \SE) = \varnothing,$  it follows that $f$ is C$^1$ on $\SU$ and $f'(x)$ is nonsingular for all $x \in \SU.$
    The inverse function theorem thus implies that, for all $x \in \SU,$ 
    there exists an open neighborhood $U_x\subseteq\SU$ of $x$ and $V_x\subset\BBR^n$ of $f(x)$ such that $V_x = f(U_x),$
    $f$ is bijective on $U_x,$ and $f^{-1}$ is C$^1$ on $V_x$ \cite[Theorem 9.17]{rudin1964}, which implies that, for all $x\in\SU,$ $f \colon U_x \to V_x$ is a C$^1$ diffeomorphism.
    Note that $\cup_{x\in\SU} U_x$ is an open covering of $\SU$ and $\BBR^n$ is a Lindel$\ddot{\rmo}$f space \cite[p. 96]{jameson1974}.
    Hence, there exists a countable subset $\SJ \subset \SU$ such that $\SU \subseteq \cup_{x \in \SJ} U_x$ and thus, for all $x \in \SJ,$ $f \colon U_x \to V_x$ is a C$^1$ diffeomorphism.
    
    Next, let $\SP \subset \BBR^n$ be a measurable set such that $\mu(\SP) > 0.$
    Then, since $\mu(\SD_0) = \mu(\SE) = 0$ and $\SD_0,$ $\SE,$ and $\SU$ are disjoint, 
    \small
    \begin{equation*}
        \mu(\SP) = \mu(\SP \cap \SD_0) + \mu(\SP \cap \SE) + \sum_{x \in \SJ} \mu(\SP \cap \SU_x) =  \sum_{x \in \SJ} \mu(\SP \cap \SU_x),
    \end{equation*}
    \normalsize
    which implies that there exists $\chi \in \SJ$ such that $\mu(\SP \cap U_\chi) > 0.$
    Since, for all $x\in U_\chi,$ $f'(x)$ exists, the change of variables theorem implies 
    \small
    \begin{equation*}
        \mu(f(\SP \cap U_\chi)) = \int_{f(\SP \cap \SU_\chi)} d \mu(y) = \int_{\SP \cap \SU_\chi} |\det f'(x)| d \mu(x) > 0.
    \end{equation*}
    \normalsize
    Hence, $\mu(f(\SP))> 0.$
    Next, suppose $\mu(f^{-1}(\SM))>0.$
    Since $f(f^{-1}(\SM)) \subseteq \SM,$ it follows that
   \begin{equation*}
        0 < \mu(f(f^{-1}(\SM))) \le \mu(\SM) = 0,
    \end{equation*}
    which is a contradiction.
    Hence, $\mu(f^{-1}(\SM)) = 0.$
    Finally, induction implies that, for all $k\ge1,$ $\mu(f^{-k}(\SM))=0.$
\hfill $\square$

The following theorem, which is the central result of the paper, provides sufficient conditions under which the set of initial conditions for which the state trajectory of \eqref{xLure}, \eqref{yLure} converges has measure zero.

\begin{theo}\label{TheoConvergence1}
Assume that $I-A$ is nonsingular, $G(1) = 0,$ $\phi(0) = 0,$ $\phi$ is PWC$^1,$ $\phi'(0)$ exists, 
${\rm acc}(\SR_0) = \varnothing,$
$\spr(f'(x_\rme)) >1,$ and $f'(x_\rme)$ is nonsingular.
Then, $\mu(\{x_0 \colon \lim_{k\to\infty} x_k  \mbox{ exists}\}) = 0.$
\end{theo}

{\it Proof:}
Proposition \ref{propequilibrium}$v$) implies that $x_\rme$ is a fixed point of $f.$
Since $\spr(f'(x_\rme)) >1,$ define $\SX \isdef x_\rme + \SY,$  where $\SY$ is the proper subspace of $\BBR^n$ spanned by the generalized eigenvectors associated with the eigenvalues of $f'(x_\rme)$ whose magnitude is less than or equal to 1.  %

Since $f'(x_\rme)$ is nonsingular, the inverse function theorem implies that there exist open neighborhoods $U \subset \BBR^n$ of $x_\rme \in U$ and $V \subset \BBR^n$ of $f(x_\rme)$ such that $V = f(U),$ $f$ is bijective on $U$, and $f^{-1}$ is continuously differentiable on $V$ \cite[Theorem 9.17]{rudin1964}.
Then, the stable manifold theorem
(Theorem III.7 in \cite[pp. 65, 66]{shub1987})
implies that there exist a local $f$-invariant C$^1$ embedded disk $\SW \subset \BBR^n$ and 
a ball $\SB_{x_\rme}$ around $x_\rme$ in an adapted norm
such that 
$\SW$ is tangent to $\SX$ at $x_\rme,$ 
$f(\SW) \cap \SB_{x_\rme} \subset \SW,$
$\SW_{x_\rme} \isdef \cap_{p=0}^{\infty} f^{-p}(\SB_{x_\rme})\subset\SW,$ and,
since $\spr(f'(x_\rme)) >1,$ $\SW$ has codimension of at least 1, and thus $\mu(\SW) = 0.$
Furthermore, since $\SW_{x_\rme} \subset \SW,$ $\mu(\SW_{x_\rme}) = 0.$

Next, let $\chi_0 \in \{x_0 \colon \lim_{k \to \infty} x_k = x_\rme\},$ and note that there exists $k_\rml \ge 1$ such that, for all $k \ge k_\rml,$ $f^k(\chi_0) \in \SB_{x_\rme},$ 
which in turn implies that $f^{k_\rml} (\chi_0) \in \SW_{x_\rme}.$
This, in turn, implies that $\chi_0 \in \cup_{k = 0}^\infty f^{-k} (\SW_{x_\rme}),$ and thus $\{x_0 \colon \lim_{k \to \infty} x_k = x_\rme\} \subseteq \cup_{k = 0}^\infty f^{-k} (\SW_{x_\rme}).$
Hence, since $\mu(\SW_{x_\rme}) = 0,$ Lemma \ref{lem:finvMeas0} implies that 
\begin{align}
    \mu(\{x_0 \colon \lim_{k \to \infty} x_k = x_\rme\}) &\leq \mu\left(\bigcup_{k = 0}^\infty f^{-k} (\SW_{x_\rme})\right)\nn\\
    &= \sum_{k=0}^{\infty} \mu(f^{-k} (\SW_{x_\rme})) = 0,\nn
\end{align}
which, with Proposition \ref{PropConvergence}$ii)b),$ implies that
\begin{equation}
\mu(\{x_0 \colon \lim_{k\to\infty} x_k \mbox{ exists}\}) = 0.\tag*{\mbox{$\square$}}
\end{equation} 

\subsection{Boundedness of Solutions of the Lur'e Model}\label{subsec:boundednessLure}

The following definition will be used to obtain conditions for the boundedness of solutions of \eqref{xLure}, \eqref{yLure}.

\begin{defin} \label{def_aff_constr}
$\phi$ is {\it affinely constrained} if there exist $\alpha_\rml, \alpha_\rmh,s_\rml, s_\rmh \in \BBR$ and $\rho>0$ such that $s_\rml < s_\rmh$ and such that, for all $y\le s_\rml,$  $|\phi(y) - \alpha_\rml y|<\rho$ and, for all $y\ge s_\rmh,$  $|\phi(y) - \alpha_\rmh y|<\rho.$ %
Furthermore, $\phi$ is {\it affinely constrained by} $(\alpha_\rml,\alpha_\rmh).$
\end{defin}

\begin{figure}[h!]
    \centering
    \includegraphics[width = 0.85\columnwidth]{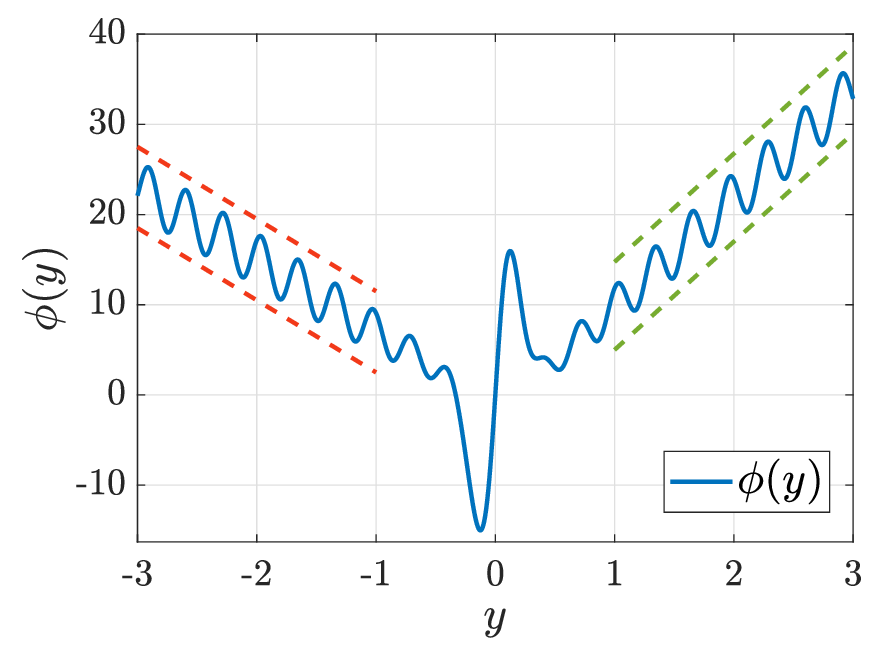}
    \caption{Plot of $\phi(y) = g(y) + h(\gamma y),$ where $g$ and $h$ are given by  \eqref{eq:ex_aff_const_g} and  \eqref{eq:ex_aff_const_h}, and $\gamma = 4, \zeta = 3,  \eta = 20, \mu = 0.125, s_\rml = -1, s_\rmh = 1.5.$
    In this case, $\phi$ is affinely constrained by $(\alpha_\rml, \alpha_\rmh),$ where $\alpha_\rml = 2 \gamma s_\rml = -8$ is the slope of the red, dashed line segments, and $\alpha_\rmh = 2 \gamma s_\rmh = 12$ is the slope of the green, dashed line segments.}
    \label{fig:Affine_Contsr_Ex}
\end{figure}

\begin{exam} \label{ex:aff_constr}
{\it
This example illustrates Definition \ref{def_aff_constr}.
}
Let $\gamma,$ $\zeta,$ $\eta,$ $\mu,$ $s_\rml,$ $s_\rmh \in \BBR$, where $\mu\ne0,$ $s_\rml < 0 < s_\rmh,$ let $\phi(y) = g(y) + h(\gamma y),$ where $g,h \colon \BBR\to\BBR$ are given by
\begin{align}
    g(y) &\isdef \zeta \tanh(y) \sin(\eta y) +  \frac{y}{\sqrt{2 \pi} \mu^3} e^{\frac{-y^2}{2 \mu^2}},\label{eq:ex_aff_const_g}\\ 
    h(y) &\isdef \begin{cases}
    s_\rml^2 + 2 s_\rml (y - s_\rml), & y \le s_\rml,\\
    y^2, & y \in (s_\rml, s_\rmh), \\
    s_\rmh^2 + 2 s_\rmh (y - s_\rmh), & y \ge s_\rmh.
    \end{cases}\label{eq:ex_aff_const_h}
\end{align}
Since $\lim_{|y|\to\infty} g(y) = 0$ it follows that $\phi$ is affinely constrained by $(2 \gamma s_\rml, 2 \gamma s_\rmh).$
Fig. \ref{fig:Affine_Contsr_Ex} shows $\phi(y)$ for all $y\in[-3, 3]$ when $\gamma = 4,$ $\zeta = 3,$ $\eta = 20,$ $\mu = 0.125,$ $s_\rml = -1,$ $s_\rmh = 1.5.$
In this case, $\phi$ is affinely constrained by $(-8, 12).$
\end{exam}

The following result provides sufficient condition under which \eqref{xLure}, \eqref{yLure} is self-excited.

\begin{theo} \label{TheoConvergence}
Assume that $I-A$ is nonsingular,
$A$ is asymptotically stable,
$G(1)=0,$  $\phi$ is continuous, and $\phi(0) = 0,$ let $\alpha_\rml, \alpha_\rmh \in \BBR,$ assume that $\phi$ is affinely constrained by $(\alpha_\rml, \alpha_\rmh),$ 
    assume that  $A_\rml \isdef A+\alpha_\rml BC$ and $A_\rmh \isdef A+\alpha_\rmh BC$ are
    asymptotically stable,
    and assume that there exists a positive-definite matrix $P \in \BBR^{n\times n}$ such that $P - A^\rmT P A,$ $P - A_\rml^\rmT P A_\rml,$ and $P - A_\rmh^\rmT P A_\rmh$ are positive definite.  
Then, the following statements hold:
\begin{enumerate}
    \item For all $x_0 \in \BBR^n,$ $(x_k)_{k=1}^{\infty}$ is bounded.
    \item  Assume that $\phi$ is PWC$^1$ and differentiable at 0, 
    ${\rm acc}(\SR_0) = \varnothing,$
    $\spr(f'(x_\rme)) >1,$ and $f'(x_\rme)$ is nonsingular.
    Then, \eqref{xLure}, \eqref{yLure} is self-excited.
\end{enumerate}
\end{theo}

{\it Proof:}
To prove $i)$,
let $s_\rml, s_\rmh \in \BBR$ and $\rho > 0$ be such that $s_\rml < s_\rmh$ and such that, for all $y\in(-\infty,s_\rml],$  $|\phi(y) - \alpha_\rml y|<\rho$ and,
for all $y\in[s_\rmh,\infty),$  $|\phi(y) - \alpha_\rmh y|<\rho.$
For all $k \ge 0,$ \eqref{xLure} can be rewritten as
\small
\begin{equation}\label{xLureEquiv0}
	x_{k+1} = \begin{cases}
		(A + \alpha_\rml BC) x_k  \\ 
		\quad  + B(\phi(C x_k)- \alpha_\rml C x_k + v), & C x_k \le s_\rml, \\
		A x_k + B(\phi(C x_k) + v), & C x_k \in (s_\rml, s_\rmh), \\
		(A + \alpha_\rmh BC) x_k  \\
		\quad + B(\phi(C x_k) - \alpha_\rmh C x_k + v), & C x_k \ge s_\rmh.
	\end{cases}
\end{equation}
\normalsize
Furthermore, defining
\begin{align*}
	A_k &\isdef \begin{cases}
		A_\rml, & C x_k \le s_\rml, \\
		A, & C x_k \in (s_\rml, s_\rmh), \\
		A_\rmh, & C x_k \ge s_\rmh,
	\end{cases} \\
	\nu_k &\isdef \begin{cases}
	\phi(C x_k) - \alpha_\rml C x_k + v, & C x_k \le s_\rml, \\
	\phi(C x_k) + v, & C x_k \in (s_\rml, s_\rmh), \\
	\phi(C x_k) - \alpha_\rmh C x_k + v, & C x_k \ge s_\rmh,
	\end{cases}
\end{align*}
\eqref{xLureEquiv0} can be written as
\begin{equation}\label{xLureEquiv}
	x_{k+1} = A_k x_k + B \nu_k.
\end{equation}
Since $\phi$ is continuous and affinely constrained by $(\alpha_\rml, \alpha_\rmh),$ it follows that $(\nu_k)_{k = 0}^{\infty}$ is bounded.
Next, 
define the positive-definite matrices 
\small
\begin{align*}
	Q_\rml \isdef P - A_\rml^\rmT P A_\rml,\quad
	Q \isdef P - A^\rmT P A,\quad
	Q_\rmh \isdef P - A_\rmh^\rmT P A_\rmh,
\end{align*}
\normalsize
and $V \colon \BBR^n \to \BBR$ such that, for all $x \in \BBR^n,$ $V(x) \isdef x^\rmT P x.$
Then, for all $k\ge0,$  \eqref{xLureEquiv} implies
\small
\begin{align*}
	&V(x_{k+1}) - V(x_k) \\
	& = \begin{cases}
	-x_k^\rmT Q_\rml x_k + 2 x_k^\rmT A_\rml^\rmT P B \nu_k + \nu_k^\rmT B^\rmT P B\nu_k, & C x_k \le s_\rml, \\
	-x_k^\rmT Q x_k + 2 x_k^\rmT A^\rmT P B \nu_k + \nu_k^\rmT B^\rmT P B\nu_k, & C x_k \in (s_\rml, s_\rmh), \\
	-x_k^\rmT Q_\rmh x_k + 2 x_k^\rmT A_\rmh^\rmT P B \nu_k + \nu_k^\rmT B^\rmT P B\nu_k, & C x_k \ge s_\rmh.
	\end{cases}
\end{align*}
\normalsize
Hence, for all $k\ge0,$
\begin{equation*}
	V(x_{k+1}) - V(x_k) \le -\gamma( \Vert x_k \Vert ) + \zeta( \Vert \nu_k \Vert ),
\end{equation*}
where $\gamma\colon[0,\infty)\to[0,\infty)$ and  $\zeta\colon [0,\infty)\to[0,\infty)$ are defined by
\begin{equation*}
	\gamma(r) \isdef \half \min(\{\lambda_{\rm min}(Q_\rml), \ \lambda_{\rm min}(Q), \ \lambda_{\rm min}(Q_\rmh)\})  r^2,
\end{equation*}	
\begin{align*}	 
	\zeta(r) &\isdef
	\bigg[ \max\left\{\tfrac{2|A_\rml^\rmT P B|^2}{\lambda_{\rm min}(Q_\rml)},  \tfrac{2|A^\rmT P B|^2}{\lambda_{\rm min}(Q)},   \tfrac{2|A_\rmh^\rmT P B|^2}{\lambda_{\rm min}(Q_\rmh)} \right\} \\ 
	   &\quad +   |B^\rmT P B|^2\bigg]  r^2.
\end{align*}
Since, for all $x\in \BBR^n,$ $\lambda_{\rm min}(P) \Vert x \Vert_2^2 \leq V(x) \leq \lambda_{\rm max}(P) \Vert x \Vert_2^2$, $\gamma$ and $\zeta$ are continuous and strictly increasing, $\gamma(0) = \zeta(0) = 0,$ and $\zeta(r) \to \infty$ as $r\to\infty,$
Lemma 3.5 of \cite{ISSCT} implies that \eqref{xLureEquiv} with input $\nu$ is input-to-state stable. 
Since $(\nu_k)_{k = 0}^{\infty}$ is bounded, it follows that, for all $x_0 \in \BBR^n,$ $(x_k)_{k=1}^{\infty}$ is bounded.

Finally, $ii)$ follows from $i)$ and Theorem \ref{TheoConvergence1}.
\hfill $\square$

    Note that Theorem \ref{TheoConvergence} assumes that the linear matrix inequality (LMI)
    \small
    \begin{equation}
        \begin{bmatrix} 
        P & 0 & 0 & 0\\ 
        0 & P - A^\rmT P A &  0 & 0 \\
        0 & 0 & P - A_\rml^\rmT P A_\rml &  0 \\ 
        0 & 0 & 0 & P - A_\rmh^\rmT P A_\rmh 
        \end{bmatrix} > 0\label{lmiineq}
    \end{equation}
    \normalsize
    is feasible, that is, there exists $P \in \BBR^{n \times n}$ such that the $4n\times 4n$ matrix in \eqref{lmiineq} is positive definite. 
    The following result provides sufficient conditions under which \eqref{lmiineq} is satisfied.
    \begin{prop}
        Assume that $\Vert A \Vert < 1,$ $\Vert A_\rml \Vert < 1,$ and $\Vert A_\rmh \Vert < 1.$ Then, \eqref{lmiineq} is satisfied with $P = I.$
    \end{prop}

    {\it Proof:} Since $\Vert A \Vert < 1,$ $\Vert A_\rml \Vert < 1,$ and $\Vert A_\rmh \Vert < 1,$ it follows that
    \begin{align*}
        I - A^\rmT A > 0, \quad
        I - A_\rml^\rmT A_\rml > 0, \quad
        I - A_\rmh^\rmT A_\rmh > 0,
    \end{align*}
    which, in turn, implies that \eqref{lmiineq} is satisfied with $P = I.$ 
    \hfill$\square$

The following is a corollary of Theorem \ref{TheoConvergence}$ii)$ when $\phi$ is bounded.
\begin{cor}\label{CorConvergence}
	Assume that $I-A$ is nonsingular, $G(1)=0,$ and $\phi(0) = 0.$ 
	Furthermore, assume that $A$ is asymptotically stable, $\phi$ is PWC$^1,$ differentiable at 0, and bounded, 
    ${\rm acc}(\SR_0) = \varnothing,$
    $\spr(f'(x_\rme)) >1,$ and $f'(x_\rme)$ is nonsingular.
	Then, \eqref{xLure}, \eqref{yLure} is self-excited.
\end{cor}

\section{Numerical examples} \label{sec:LureNumericalExamples}

Although the conditions of Theorem \ref{TheoConvergence} and Corollary \ref{CorConvergence} are not necessary, the numerical examples in this section show that, when some of these conditions are not met, the response of \eqref{xLure}, \eqref{yLure} may yield a convergent or divergent response for a nonnegligible set of initial conditions.
Examples \ref{ex_lure_b} to \ref{ex_lure_ub_3} concern cases in which some of these conditions are not met. 
Table \ref{example_tab} summarizes these examples and their objectives.
In these examples, the feasibility of the LMI in \eqref{lmiineq} is determined by using the Matlab function {\it feasp}, which is also used to compute a feasible solution when it exists.

\begin{table}[h]
\caption{Summary of Numerical Examples}
\label{example_tab}
\centering
\renewcommand{\arraystretch}{1.2}
\resizebox{\columnwidth}{!}{%
\begin{tabular}{|c|c|c|}
\hline
\textbf{Example} & \textbf{Nonlinearity type} & \textbf{Objective} \\
\hline
\ref{ex_lure_b}  & Bounded, C$^1$ & $\begin{array} {c} \mbox{Shows that Corollary \ref{CorConvergence} is false if } G(1) = 0 \mbox{ is omitted} \end{array}$ \\
\hline
\ref{ex_lure_ub}  & $\begin{array}{c} \mbox{Unbounded,} \ \rmC^1,\\ \mbox{affinely constrained by } (\alpha, \alpha) \end{array}$ & $\begin{array}{c} \mbox{Shows that Theorem \ref{TheoConvergence} is false if either}\\ \spr(A+\alpha BC)<1   \mbox{ or } \spr(f'(x_\rme)) > 1 \mbox{ is omitted} \end{array}$ \\
\hline
\ref{ex_lure_ub_2}  & $\begin{array}{c} \mbox{Unbounded,} \ {\rm PWC}^1,\\ \mbox{affinely constrained by }(\alpha_\rml, \alpha_\rmh) \end{array}$ & $\begin{array}{c} \mbox{Shows that Theorem \ref{TheoConvergence} is false if either} \\ \spr(A+ \alpha_\rml BC)<1  \mbox{ or } \spr(A+ \alpha_\rmh BC)<1 \mbox{ is omitted} \end{array}$ \\
\hline
\ref{ex_lure_ub_3}  & $\begin{array}{c} \mbox{Unbounded,} \ {\rm PWC}^1,\\ \mbox{affinely constrained by }(\alpha_\rml, \alpha_\rmh) \end{array}$ & $\begin{array}{c} \mbox{Shows that Theorem \ref{TheoConvergence} is false if}\\  \mbox{the feasibility of \eqref{lmiineq} is omitted} \end{array}$ \\
\hline
\end{tabular}
}
\end{table}

\begin{exam}\label{ex_lure_b}
{\it
This example shows that Corollary \ref{CorConvergence} is false if the assumption that $G(1) = 0$ is omitted.
}
Let $v = 5,$ $\phi(y) = \tanh(y),$ and $G (\bfz) = -1/(\bfz^2 - \bfz + 0.5)$ with minimal realization
\begin{equation*}
     A = \begin{bmatrix} 1 & -0.5 \\ 1 & 0 \end{bmatrix}, \ B = \begin{bmatrix} 1 \\ 0 \end{bmatrix}, \ C = \begin{bmatrix} 0 & -1\end{bmatrix}.
\end{equation*}
Note that $\phi$ is $C^1,$ 
bounded, and $\phi(0) = 0.$
Root-locus properties imply that $A+\phi'(y) BC$ is singular if and only if $\phi'(y) = -0.5.$
Since, for all $y \in \BBR,$ $\phi'(y) = \sech^2(y) \in [0, 1],$  it follows that $A+\phi'(y) BC$ is nonsingular, and thus 
${\rm acc}(\SR_0) = \varnothing.$
Furthermore, $I - A$ is nonsingular, $A$ is asymptotically stable, and $\spr(f'(x_\rme))>1.$
Since $G(1)\ne0,$ it follows that the assumptions of Corollary \ref{CorConvergence} are not satisfied.
Accordingly, Fig. \ref{fig:ex_lure_b_G1} shows that, for the indicated initial states, the response of \eqref{xLure}, \eqref{yLure} converges.

Next, let $G (\bfz) = -(\bfz - 1)/(\bfz^2 - \bfz + 0.5)$ with minimal realization
\begin{equation*}
    A = \begin{bmatrix} 1 & -0.5 \\ 1 & 0 \end{bmatrix}, \ B = \begin{bmatrix} 1 \\ 0 \end{bmatrix}, \ C = \begin{bmatrix} 1 & -1\end{bmatrix}.
\end{equation*}
Root-locus properties imply that $A+\phi'(y) BC$ is singular if and only if $\phi'(y) = -0.5.$
Since, for all $y \in \BBR,$ $\phi'(y) = \sech^2(y) \in [0, 1],$ it follows that $A+\phi'(y) BC$ is nonsingular, that is, $\SR_0 = \varnothing.$
Furthermore, $I - A$ is nonsingular, $A$ is asymptotically stable, and $\spr(f'(x_\rme)) > 1.$
Since $G(1) = 0,$ all of the assumptions of Corollary \ref{CorConvergence} are satisfied.
Accordingly, Fig. \ref{fig:ex_lure_b_G2} shows that, for the indicated initial states except the equilibrium, the response of \eqref{xLure}, \eqref{yLure} does not converge and is bounded.
\hfill$\huge\Diamond$
\begin{figure}[h]
    \centering
    \includegraphics[width = \columnwidth]{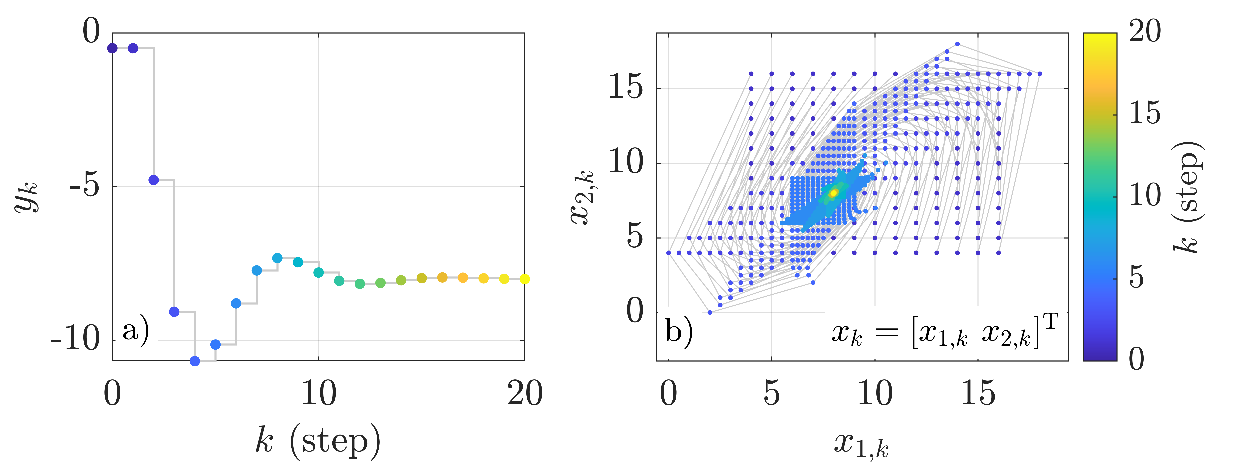}
    \caption{Example \ref{ex_lure_b}: Response of \eqref{xLure}, \eqref{yLure} for  $G (\bfz) = \frac{-1}{\bfz^2 - \bfz + 0.5},$ $v = 5,$ and $\phi(y) = \tanh(y).$
    For all $k \in [0, 20],$ $a)$ shows $y_k$ for $x_0 =  [0.5 \ \ 0.5]^\rmT.$ 
    For all $k \in [0, 20],$ $b)$ shows $x_k$ for all $x_0 \in \{4, 5, \ldots, 16\} \times \{4, 5, \ldots, 16\}.$
    The gray lines follow the trajectory from each initial state.
    Note that all state trajectories converge to $x = [8 \ \ 8]^\rmT,$ which is an asymptotically stable equilibrium.
    }
    \label{fig:ex_lure_b_G1}
\end{figure}
\begin{figure}[h]
    \centering
    \includegraphics[width = \columnwidth]{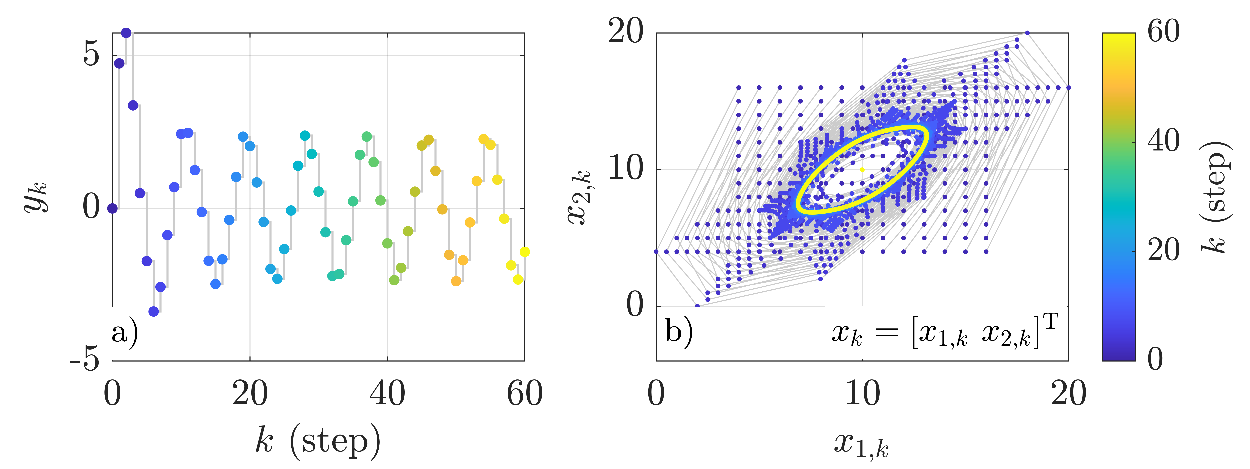}
    \caption{Example \ref{ex_lure_b}:  Response of \eqref{xLure}, \eqref{yLure} for $G (\bfz) = \frac{-(\bfz - 1)}{\bfz^2 - \bfz + 0.5},$ $v = 5,$ and $\phi(y) = \tanh(y).$
    For all $k \in [0, 60],$ $a)$ shows $y_k$ for $x_0 =  [0.5 \ \ 0.5]^\rmT.$ 
    For all $k \in [0, 60],$ $b)$ shows $x_k$ for all $x_0 \in \{4, 5, \ldots, 16\} \times \{4, 5, \ldots, 16\}.$
    The gray lines follow the trajectory from each initial state.
    Note that each state trajectory is bounded and does not converge, except for the state trajectory for $x_0 = [10 \ \ 10]^\rmT = x_\rme,$ which is an unstable equilibrium.
    }
    \label{fig:ex_lure_b_G2}
\end{figure}
\end{exam}

\begin{exam}\label{ex_lure_ub}
{\it
This example shows that Theorem \ref{TheoConvergence} is false if either $\spr(A+\alpha BC)<1$   or $\spr(f'(x_\rme)) > 1$ is omitted.
}
Let $v = 5,$ $\alpha,$ $\beta \in \BBR,$ where $\beta \ne 0,$ $\phi(y) = \alpha y + \beta \sin(y),$ and $G (\bfz) = (\bfz - 1)/(\bfz^2 - \bfz + 0.5)$ with minimal realization
\begin{equation*}
    A = \begin{bmatrix} 1 & -0.5 \\ 1 & 0 \end{bmatrix}, \ B = \begin{bmatrix} 1 \\ 0 \end{bmatrix}, \ C = \begin{bmatrix} 1 & -1 \end{bmatrix}.
\end{equation*}
Note that $\phi$ is C$^1$ 
and affinely constrained by $(\alpha, \alpha)$ since, for all $y\in\BBR,$ $|\phi(y) - \alpha y| = |\beta\sin(y)| \le |\beta|.$
Next, root-locus properties imply that $A + \phi'(y) BC$ is singular if and only if $\phi'(y) = -0.5$
Then, since $\phi' (y) = \alpha + \beta \cos(y)$, $\SR_0 = \{y\in \BBR \colon \cos(y) = (-0.5 - \alpha) / \beta\}$ is countable and thus
${\rm acc}(\SR_0) = \varnothing.$
Furthermore, $I-A$ is nonsingular, $A$ is asymptotically stable, $G(1) = 0,$ and $\phi(0) = 0.$
In particular, for $\alpha = 0.25$ and $\beta = 0.05,$ it follows that $\spr(A+\alpha BC) <1$ and $\spr(f'(x_\rme))<1.$
Hence, the assumptions of Theorem \ref{TheoConvergence} are not satisfied.
Accordingly, Fig. \ref{fig:ex_lure_ub_AS} shows that, for the indicated initial states, the response of \eqref{xLure}, \eqref{yLure} converges.

Furthermore, for $\alpha = 0.75$ and $\beta = 0.5,$ it follows that $\spr(A+\alpha BC) >1$ and $\spr(f'(x_\rme))>1.$
Hence, the assumptions of Theorem \ref{TheoConvergence} are not satisfied.
Accordingly, Fig. \ref{fig:ex_lure_ub_US} shows that, for the indicated initial states except the equilibrium, the response of \eqref{xLure}, \eqref{yLure} diverges.

Finally, for $\alpha = 0.25$ and $\beta = 0.5,$ it follows that $\spr(A+\alpha BC) <1$ and $\spr(f'(x_\rme))>1.$
Furthermore, \eqref{lmiineq} is feasible with
\begin{equation*}
P = \begin{bmatrix} 2.24 & -1.32 \\ -1.32 & 1.62 \end{bmatrix}.
\end{equation*}
Hence, the assumptions of Theorem \ref{TheoConvergence} are satisfied.
Accordingly, Fig. \ref{fig:ex_lure_ub_SES} shows that, for the indicated initial states except the equilibrium, the response of \eqref{xLure}, \eqref{yLure} does not converge and is bounded.
\hfill$\huge\Diamond$
\begin{figure}[h]
    \centering
    \includegraphics[width = \columnwidth]{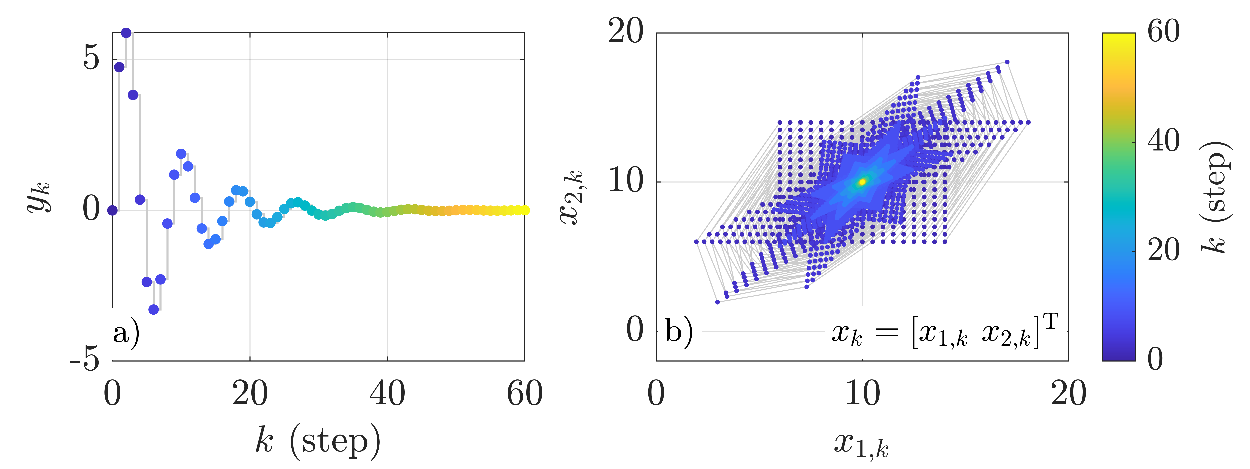}
    \caption{Example \ref{ex_lure_ub}: Response of \eqref{xLure}, \eqref{yLure} for $G (\bfz) = (\bfz - 1)/(\bfz^2 - \bfz + 0.5),$ $v = 5,$ $\phi(y) = \alpha y + \beta \sin(y),$ and $\alpha = 0.25,$ $\beta = 0.05.$
    For all $k \in [0, 60],$ $a)$ shows $y_k$ for $x_0 =  [0.5 \ \ 0.5]^\rmT.$ 
    For all $k \in [0, 60],$ $b)$ shows $x_k$ for all $x_0 \in \{6, 6.5, \ldots, 14\} \times \{6, 6.5, \ldots, 14\}.$  
    The gray lines follow the trajectory from each initial state.
    Note that all state trajectories converge to $x = [10 \ \ 10]^\rmT,$ which is an asymptotically stable equilibrium.}
    \label{fig:ex_lure_ub_AS}
\end{figure}
\begin{figure}[h]
    \centering
    \includegraphics[width = \columnwidth]{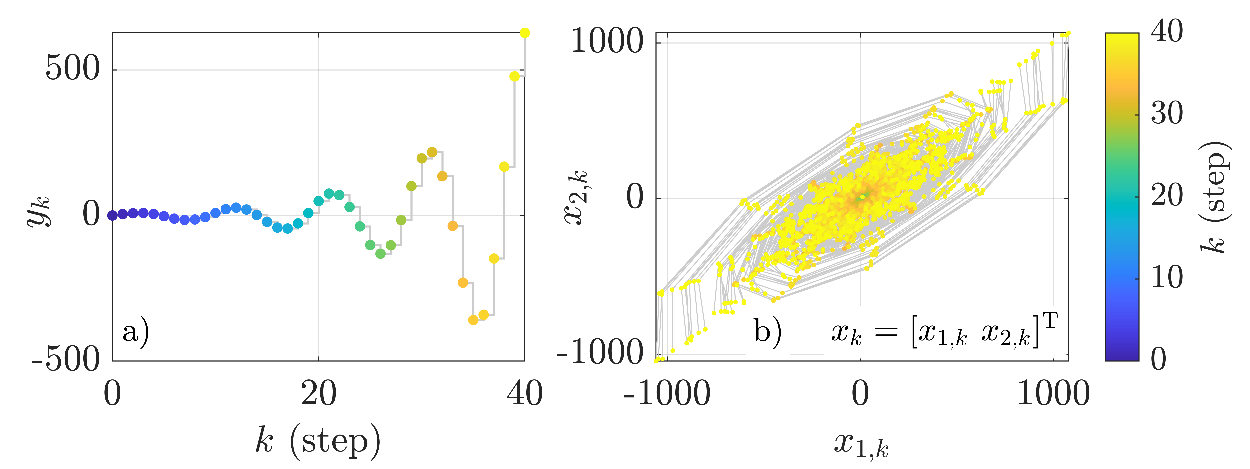}
    \caption{Example \ref{ex_lure_ub}:  Response of \eqref{xLure}, \eqref{yLure} for $G (\bfz) = (\bfz - 1)/(\bfz^2 - \bfz + 0.5),$ $v = 5,$ $\phi(y) = \alpha y + \beta \sin(y),$ and $\alpha = 0.75,$ $\beta = 0.5.$
    For all $k \in [0, 40],$ $a)$ shows $y_k$ for $x_0 =  [0.5 \ \ 0.5]^\rmT.$ 
    For all $k \in [0, 40],$ $b)$ shows $x_k$ for all $x_0 \in \{6, 6.5, \ldots, 14\} \times \{6, 6.5, \ldots, 14\}.$  
    The gray lines follow the trajectory from each initial state.
    Note that all state trajectories diverge, except for the state trajectory with $x_0 = [10 \ \ 10]^\rmT = x_\rme,$ which is an unstable equilibrium.}
    \label{fig:ex_lure_ub_US}
\end{figure}
\begin{figure}[h]
    \centering
    \includegraphics[width = \columnwidth]{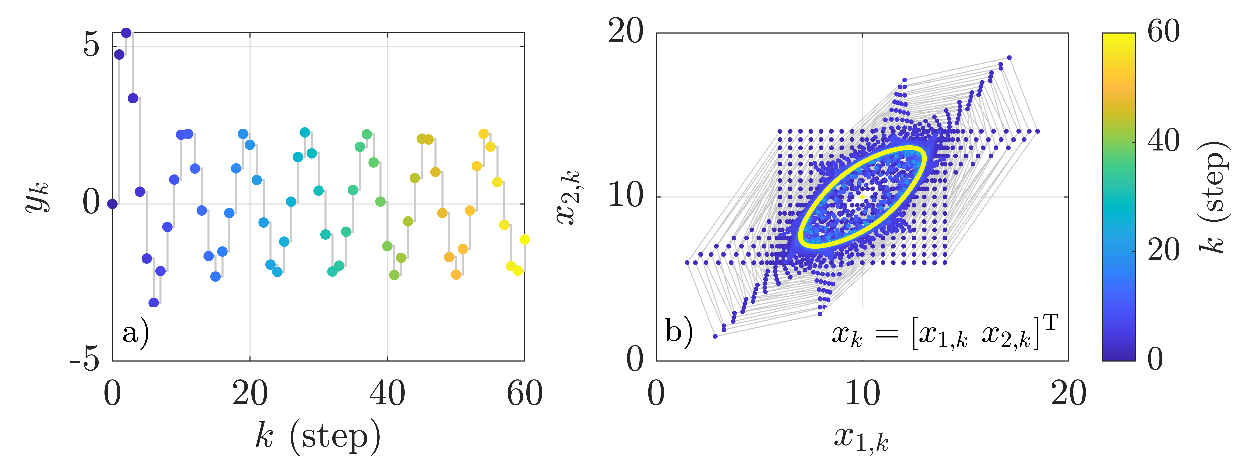}
    \caption{Example \ref{ex_lure_ub}:  Response of \eqref{xLure}, \eqref{yLure} for $G (\bfz) = (\bfz - 1)/(\bfz^2 - \bfz + 0.5),$ $v = 5,$ $\phi(y) = \alpha y + \beta \sin(y),$ and $\alpha = 0.25,$ $\beta = 0.5.$
    For all $k \in [0, 60],$ $a)$ shows $y_k$ for $x_0 =  [0.5 \ \ 0.5]^\rmT.$ 
    For all $k \in [0, 60],$ $b)$ shows $x_k$ for all $x_0 \in \{6, 6.5, \ldots, 14\} \times \{6, 6.5, \ldots, 14\}.$  
    The gray lines follow the trajectory from each initial state.
    Note that each state trajectory is bounded and does not converge, except for the state trajectory for $x_0 = [10 \ \ 10]^\rmT = x_\rme,$ which is an unstable equilibrium.}
    \label{fig:ex_lure_ub_SES}
\end{figure}
\end{exam}

\begin{exam}\label{ex_lure_ub_2}
{\it
This example shows that Theorem \ref{TheoConvergence} is false if either $\spr(A+ \alpha_\rml BC)<1$ or $\spr(A+ \alpha_\rmh BC)<1$ is omitted.
}
Let $v=5,$ let $\mu, s_\rml,$ $s_\rmh \in \BBR$, where $\mu\ne0,$ $s_\rml < 0 < s_\rmh,$ let $\phi(y) = g (y) + h(y),$ where $g, h \colon \BBR\to\BBR$ are given by
\begin{align}
    g (y) &\isdef \frac{y}{\sqrt{2 \pi} \mu^3} e^{\frac{-y^2}{2 \mu^2}}, \label{ex_lure_ub_2_NL_f} \\ 
    h (y) &\isdef \begin{cases}
    s_\rml^2 + s_\rml (y - s_\rml), & y \le s_\rml,\\
    y^2, & y \in (s_\rml, s_\rmh), \\
    s_\rmh^2 + s_\rmh (y - s_\rmh), & y \ge s_\rmh,
    \end{cases} \label{ex_lure_ub_2_NL_g}
\end{align}
and let $G(\bfz) = \tfrac{\bfz (\bfz - 1)}{\bfz^3 - 0.5 \bfz^2 + 0.25}$ with minimal realization
\small
\begin{equation*}
    A = \begin{bmatrix} 0.5 & 0 & -0.25 \\ 1 & 0 & 0 \\ 0 & 1 & 0 \end{bmatrix}, \ B =  \begin{bmatrix} 1 \\ 0 \\ 0 \end{bmatrix}, \ C = \begin{bmatrix} 1 & -1 & 0 \end{bmatrix}.
\end{equation*}
\normalsize
Note that $\phi$ is not C$^1$ but it is PWC$^1$ with $\SSS = \{s_\rml, s_\rmh\}$ and, since $\lim_{|y|\to\infty} g (y) = 0,$ $\phi$ is affinely constrained by $(s_\rml, s_\rmh).$
Next, since $G(0) = 0,$ root-locus properties imply that, for all $y\in\SR,$ $A + \phi'(y) BC$ is nonsingular, and thus
${\rm acc}(\SR_0) = \varnothing.$
Furthermore, $I - A$ is nonsingular, $A$ is asymptotically stable, $G(1) = 0,$ and $\phi(0) = 0.$

In particular, for $\mu = 0.5,$ $s_\rml = -2,$ and $s_\rmh = 0.2,$ it follows that $\spr(A+s_\rml BC) > 1,$ $\spr(A+s_\rmh BC) <1,$ and, since $\phi'(0) = g'(0) = 3.19,$ $\spr(f'(x_\rme))>1.$
Hence, the assumptions of Theorem \ref{TheoConvergence} are not satisfied.
Accordingly, Fig. \ref{fig:ex_lure_ub_2_US} shows that, for some initial states, the response of \eqref{xLure}, \eqref{yLure} is unbounded.

Furthermore, for $\mu = 0.5,$ $s_\rml = -0.4,$ and $s_\rmh = 0.2,$ it follows that $\spr(A+s_\rml BC) < 1,$ $\spr(A+s_\rmh BC) <1,$ and, since $\phi'(0) =  3.19,$ $\spr(f'(x_\rme))>1.$
Furthermore, \eqref{lmiineq} is feasible with
\begin{equation*}
P = \begin{bmatrix} 105. 65 & -20.67 & -7.47 \\ -20.67 & 68.99 & -6.21 \\ -7.47 & -6.21 & 34.77 \end{bmatrix}.
\end{equation*}
Hence, the assumptions of Theorem \ref{TheoConvergence} are satisfied.
Accordingly, Fig. \ref{fig:ex_lure_ub_2_US} shows that, for the indicated initial states, the response of \eqref{xLure}, \eqref{yLure} is bounded and does not converge.
\hfill$\huge\Diamond$
\begin{figure}[h]
    \centering
    \includegraphics[width = \columnwidth]{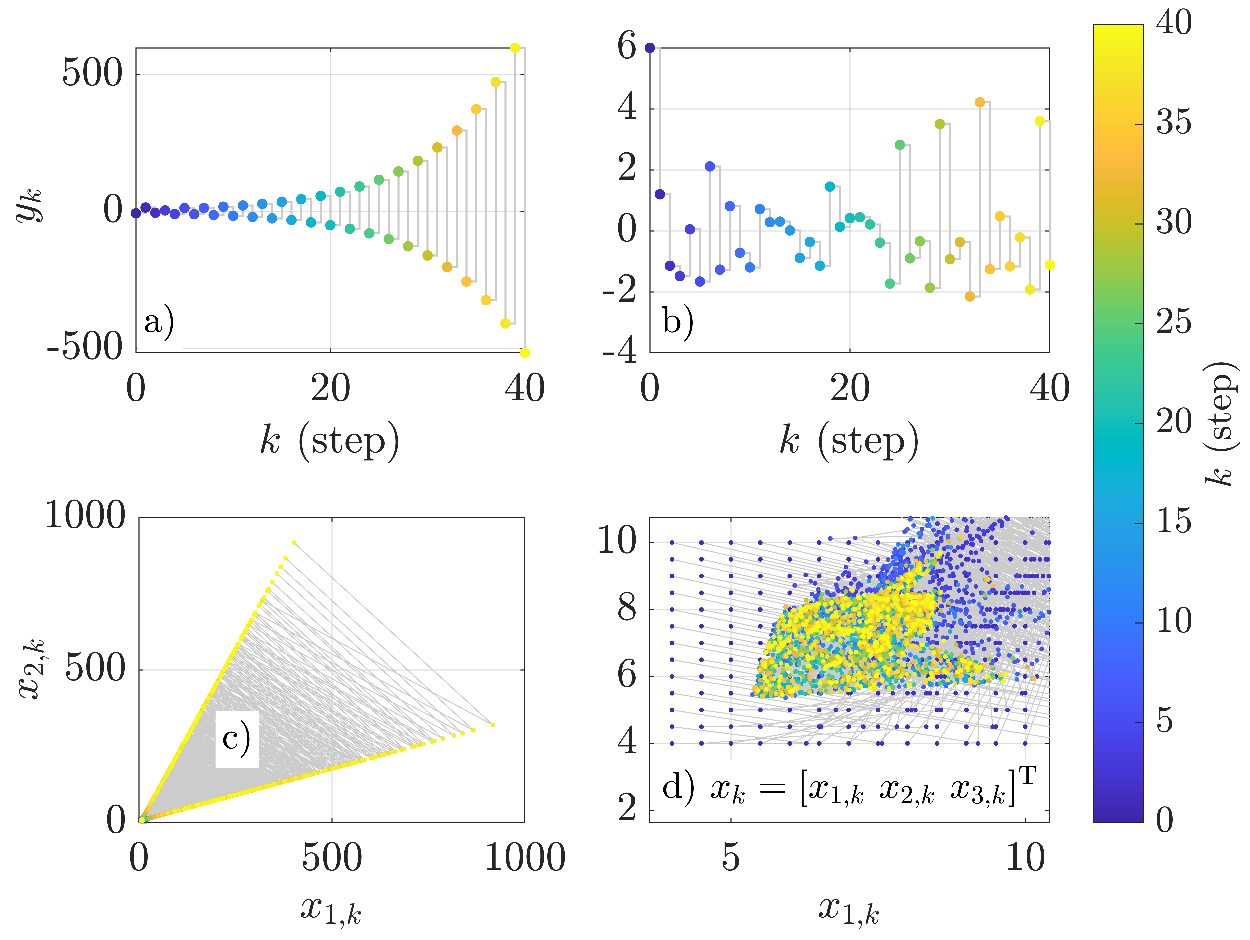}
    \caption{Example \ref{ex_lure_ub_2}: Response of \eqref{xLure}, \eqref{yLure} for $G(\bfz) = \tfrac{\bfz (\bfz - 1)}{\bfz^3 - 0.5 \bfz^2 + 0.25},$ $v = 5,$ $\phi(y) = g(y) + h(y),$ where $g$ and $h$ are given by \eqref{ex_lure_ub_2_NL_f} and \eqref{ex_lure_ub_2_NL_g}, and $\mu = 0.5, s_\rml = -2, s_\rmh = 0.2.$
    For all $k \in [0, 40],$ $a)$ shows $y_k$ for $x_0 =  [4 \ \ 10 \ \ 0]^\rmT.$ 
    For all $k \in [0, 40],$ $b)$ shows $y_k$ for $x_0 =  [10 \ \ 4 \ \ 0]^\rmT.$ 
    For all $k \in [0, 40],$ $c)$ shows $x_k$ for all $x_0 \in \{4, 5, \ldots, 10\} \times \{4, 5, \ldots, 10\} \times \{0\}.$  
    $d)$ is a magnified version of $c).$
    The gray lines follow the trajectory from each initial state.
    Note that, while some state trajectories remain bounded, the response of \eqref{xLure}, \eqref{yLure} is unbounded for some initial states.}
    \label{fig:ex_lure_ub_2_US}
\end{figure}
\begin{figure}[h]
    \centering
    \includegraphics[width = \columnwidth]{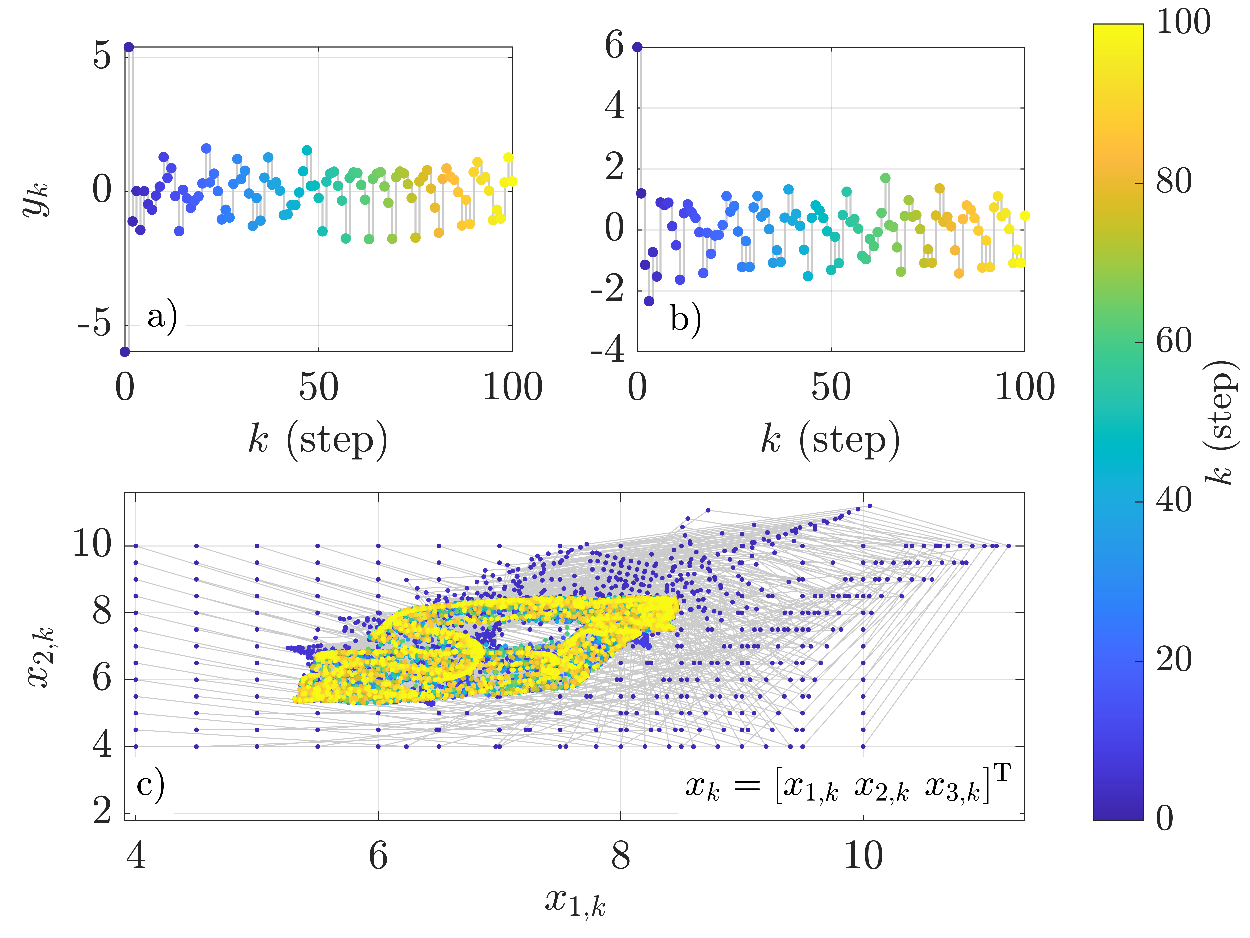}
    \caption{Example \ref{ex_lure_ub_2}:  Response of \eqref{xLure}, \eqref{yLure} for $G(\bfz) = \tfrac{\bfz (\bfz - 1)}{\bfz^3 - 0.5 \bfz^2 + 0.25},$ $v = 5,$ $\phi(y) = g(y) + h(y),$ where $g$ and $h$ are given by \eqref{ex_lure_ub_2_NL_f} and \eqref{ex_lure_ub_2_NL_g}, and $\mu = 0.5, s_\rml = -0.4, s_\rmh = 0.2.$
    For all $k \in [0, 100],$ $a)$ shows $y_k$ for $x_0 =  [4 \ \ 10 \ \ 0]^\rmT.$ 
    For all $k \in [0, 100],$ $b)$ shows $y_k$ for $x_0 =  [10 \ \ 4 \ \ 0]^\rmT.$ 
    For all $k \in [0, 100],$ $c)$ shows $x_k$ for all $x_0 \in \{4, 5, \ldots, 10\} \times \{4, 5, \ldots, 10\} \times \{0\}.$  
    The gray lines follow the trajectory from each initial state.
    Note that each state trajectory is bounded and does not converge.}
    \label{fig:ex_lure_ub_2_SES}
\end{figure}
\end{exam}

\begin{exam}\label{ex_lure_ub_3}
{\it
This example shows that Theorem \ref{TheoConvergence} is false if the assumption that \eqref{lmiineq} is feasible is omitted.
}
Let $v=5,$ let $\gamma,$ $\mu,$ $\eta,$ $s_\rml,$ $ s_\rmh \in \BBR$, where $\mu,$ $ \eta$ are nonzero and $s_\rml < 0 < s_\rmh,$ let $\phi$ be given by
\footnotesize
\begin{equation}
    \phi(y) = \begin{cases}
    s_\rml (s_\rml^2 + \gamma) + 3 s_\rml^2 (y - s_\rml) + \mu \sin(\eta(y - s_\rml)), & y \le s_\rml,\\
    y^3 + \gamma y, & y \in (s_\rml, s_\rmh), \\
    s_\rml (s_\rmh^2 + \gamma) + 3 s_\rmh^2 (y - s_\rmh) + \mu \sin(\eta(y - s_\rmh)), & y \ge s_\rmh,
    \end{cases} \label{ex_lure_ub_3_NL_phi}
\end{equation}
\normalsize
and let
\begin{equation} \label{ex_lure_ub_G_3}
G(\bfz) = \tfrac{\bfz^3 - 1.1 \bfz^2 + 0.88 \bfz - 0.78}{\bfz^4 + 0.1 \bfz^3 + 0.77 \bfz^2 - 10^{-3} \bfz - 7.8 \cdot 10^{-3}}
\end{equation}
with minimal realization
%
%
\begin{align*}
    A &= \begin{bmatrix} -0.1 & -0.77 & 10^{-3} & 7.8\cdot 10^{-3}  \\ 1 & 0 & 0 & 0 \\ 0 & 1 & 0 & 0 \\ 0 & 0 & 1 & 0 \end{bmatrix}, \ B =  \begin{bmatrix} 1 \\ 0 \\ 0 \\ 0 \end{bmatrix},\\
    C &= \begin{bmatrix} 1 & -1.1 & 0.88 & -0.78 \end{bmatrix}.
\end{align*}
%
%
Note that $\phi$ is not C$^1$ but it is PWC$^1$ with $\SSS = \{s_\rml, s_\rmh\},$ for all $y \le s_\rml,$ $|\phi(y) - 3 s_\rml^2 y| = |\mu \sin(\eta(y - s_\rml)) - 2 s_\rml^3| \le |\mu| + 2 |s_\rml|^3,$ and, for all $y \ge s_\rmh,$ $|\phi(y) - 3 s_\rmh^2 y| = |\mu \sin(\eta(y - s_\rml))- 2 s_\rmh^3| \le |\mu| + 2 |s_\rmh|^3.$
Hence, $\phi$ is affinely constrained by $(3 s_\rml^2, 3 s_\rmh^2).$
Next, root-locus properties imply that $A + \phi'(y) BC$ is singular if and only if $\phi'(y) = 0.01.$
For all $y\in \SR,$ $\phi'$ is given by
\begin{equation*}
    \phi'(y) = \begin{cases}
    3 s_\rml^2 + \mu \eta \cos(\eta(y - s_\rml)), & y < s_\rml,\\
    3 y^2 + \gamma, & y \in (s_\rml, s_\rmh), \\
    3 s_\rmh^2 + \mu \eta \cos(\eta(y - s_\rmh)), & y > s_\rmh,
    \end{cases}
\end{equation*}
which implies that 
\small
\begin{align*}
\SR_0 \subset & \{- \sqrt{|0.01 - \gamma| / 3}, \sqrt{|0.01 - \gamma| / 3}\} \\ 
&\cup \{y \in \SR \colon y < s_\rml \mbox{ and } 0.01 - 3 s_\rml^2 = \mu \eta \cos(\eta(y - s_\rml))\} \\
&\cup \{y \in \SR \colon y > s_\rmh \mbox{ and } 0.01 - 3 s_\rmh^2 = \mu \eta \cos(\eta(y - s_\rmh))\},
\end{align*}
\normalsize
which in turn implies that $\SR_0$ is countable and thus
${\rm acc}(\SR_0) = \varnothing.$
Furthermore, $I-A$ is nonsingular, $A$ is asymptotically stable, $G(1) = 0,$ and $\phi(0) = 0.$

In particular, for $\gamma = 1.5,$ $\mu = 0.1,$ $\eta = 40,$ $s_\rml = -0.29,$ $s_\rmh = 0.62,$ it follows that $\spr(A + 3 s_\rml^2 BC) < 1,$ $\spr(A + 3 s_\rmh^2 BC) < 1,$ and $\spr(f'(x_\rme)) > 1.$ 
However, \eqref{lmiineq} in infeasible.
Hence, the assumptions of Theorem \ref{TheoConvergence} are not satisfied.
Accordingly, Fig. \ref{fig:ex_lure_ub_3_US} shows that the response of \eqref{xLure}, \eqref{yLure} is unbounded for some initial states.

Furthermore, for $\gamma = 1.5,$ $\mu = 0.1,$ $\eta = 40,$ $s_\rml = -0.29,$ $s_\rmh = 0.29,$ it follows that $\spr(A + 3 s_\rml^2 BC) < 1,$ $\spr(A + 3 s_\rmh^2 BC) < 1,$ and $\spr(f'(x_\rme)) > 1.$
Furthermore, \eqref{lmiineq} is feasible with
\footnotesize	
\begin{equation*}
    P = \begin{bmatrix} 2.34  & -1.05 \cdot 10^{-1}  &  1.14  &  -1.13 \cdot 10^{-1} \\
   -1.04 \cdot 10^{-1}  &  1.74  & -1.07 \cdot 10^{-1}  &  6.35 \cdot 10^{-1} \\
    1.14  & -1.07 \cdot 10^{-1}  &  1.21  & -3.58 \cdot 10^{-2} \\
  -1.13 \cdot 10^{-1}  &  6.35 \cdot 10^{-1}  & -3.58 \cdot 10^{-2}  &  6.10 \cdot 10^{-1} \end{bmatrix}.
\end{equation*}
\normalsize
Hence, the assumptions of Theorem \ref{TheoConvergence} are satisfied.
Accordingly, Fig. \ref{fig:ex_lure_ub_3_SES} shows that, for the indicated initial states, the response of \eqref{xLure}, \eqref{yLure} is bounded and does not converge.
\hfill$\huge\Diamond$
\begin{figure}[h]
    \centering
    \includegraphics[width = \columnwidth]{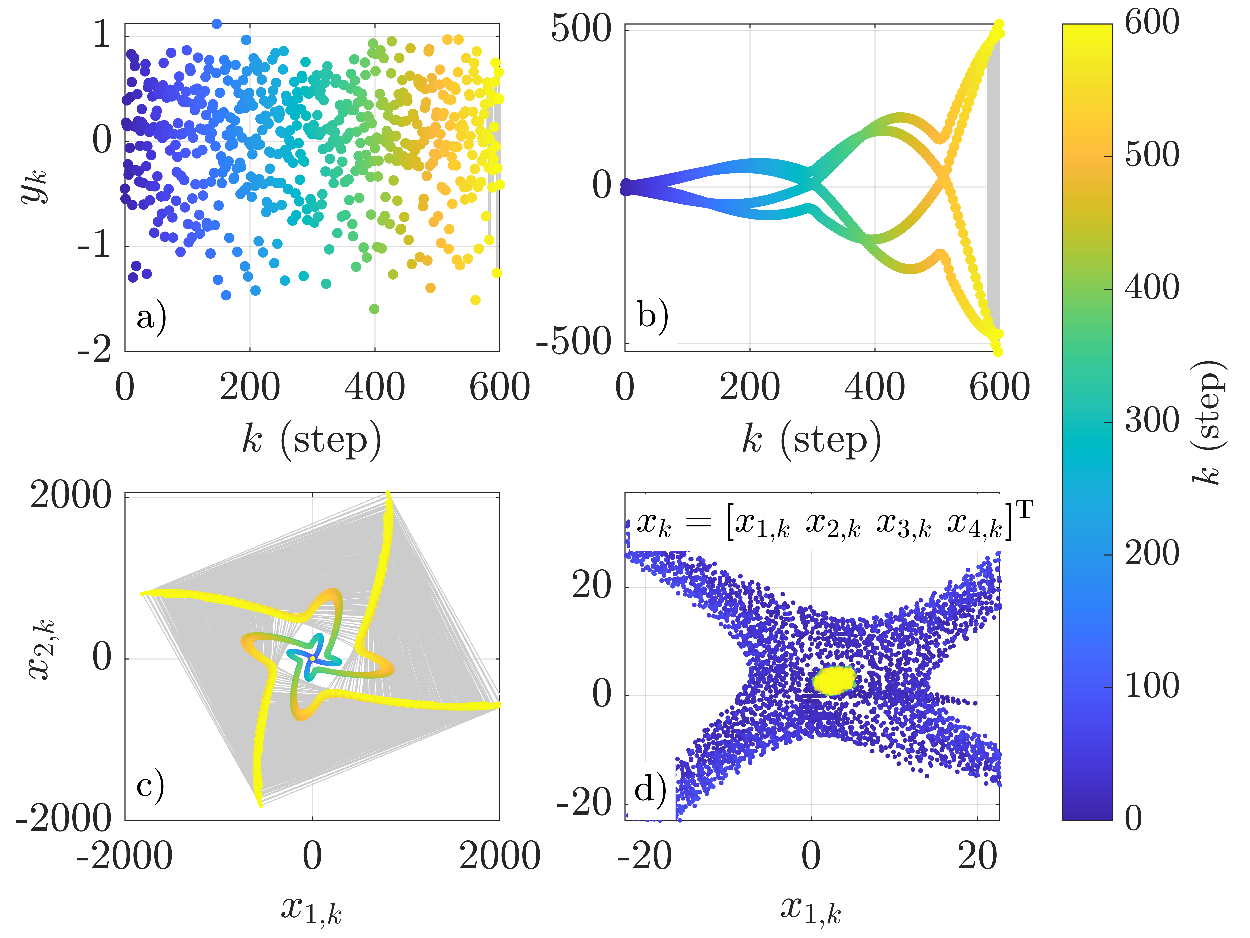}
    \caption{Example \ref{ex_lure_ub_3}: Response of \eqref{xLure}, \eqref{yLure} for $G$ given by \eqref{ex_lure_ub_G_3}, $v = 5,$ $\phi$ is given by \eqref{ex_lure_ub_3_NL_phi}, and $\gamma = 1.5,$ $\mu = 0.1,$ $\eta = 40,$ $s_\rml = -0.29,$ $s_\rmh = 0.62.$
    For all $k \in [0, 600],$ $a)$ shows $y_k$ for $x_0 =  [2 \ \ 4 \ \ 4 \ \ 2]^\rmT.$ 
    For all $k \in [0, 600],$ $b)$ shows $y_k$ for $x_0 =  [-2 \ \ 4 \ \ -4 \ \ 2]^\rmT.$ 
    For all $k \in [0, 600],$ $c)$ shows $x_k$ for all $x_0 \in \{-4, -3, \ldots, 4\} \times \{4\} \times \{-4, -3, \ldots, 4\} \times \{2\}.$  
    $d)$ is a magnified version of $c).$
    For all $k \in [580, 600],$ the gray lines follow the trajectory from each initial state. 
    Note that, while some state trajectories remain bounded, the response of \eqref{xLure}, \eqref{yLure} is unbounded for some initial states.
    }
    \label{fig:ex_lure_ub_3_US}
\end{figure}
\begin{figure}[h]
    \centering
    \includegraphics[width = \columnwidth]{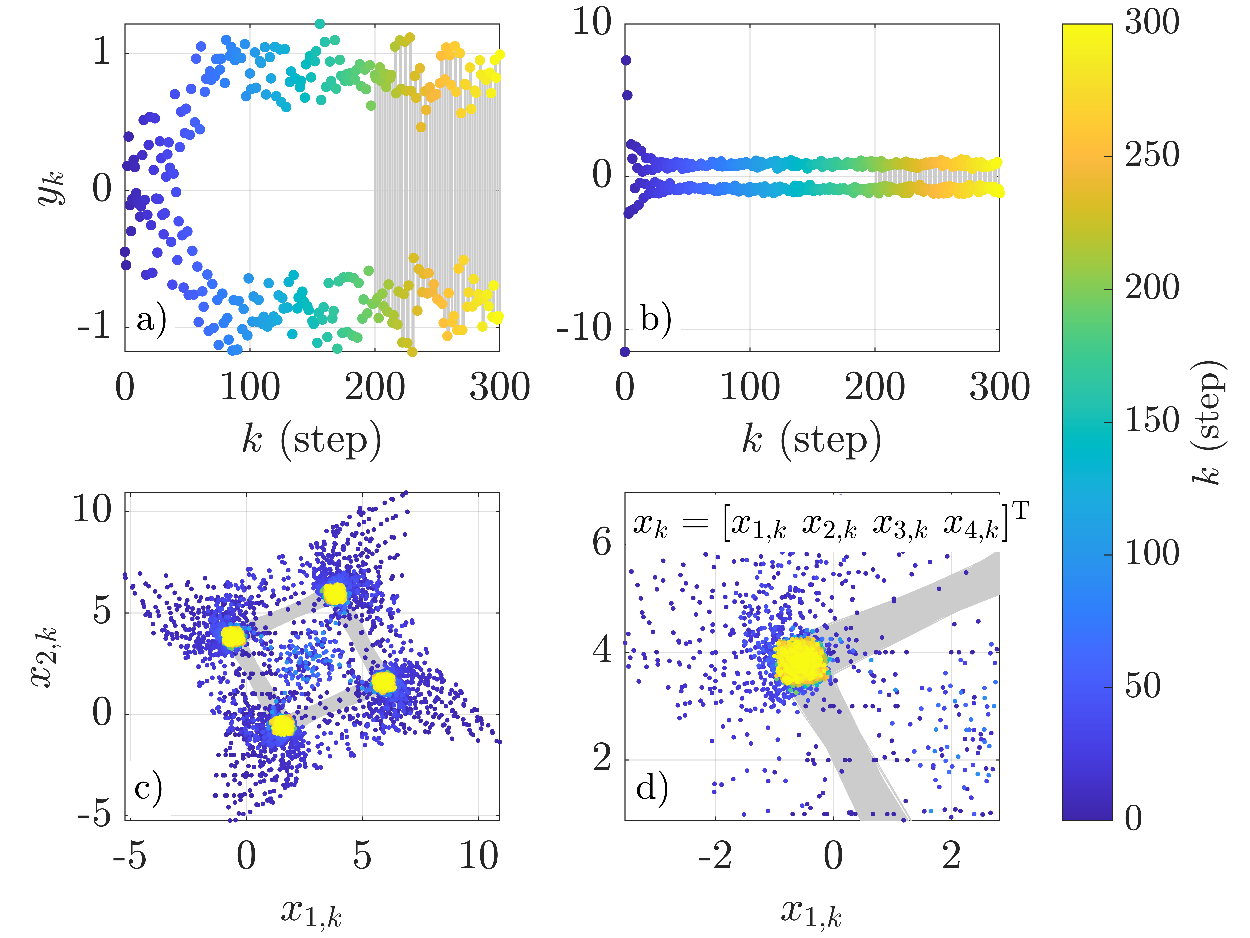}
    \caption{Example \ref{ex_lure_ub_3}:  Response of \eqref{xLure}, \eqref{yLure} for $G$ given by \eqref{ex_lure_ub_G_3}, $v = 5,$ $\phi$ is given by \eqref{ex_lure_ub_3_NL_phi}, and $\gamma = 1.5,$ $\mu = 0.1,$ $\eta = 40,$ $s_\rml = -0.29,$ $s_\rmh = 0.29.$
    For all $k \in [0, 300],$ $a)$ shows $y_k$ for $x_0 =   [2 \ \ 4 \ \ 4 \ \ 2]^\rmT.$ 
    For all $k \in [0, 300],$ $b)$ shows $y_k$ for $x_0 =  [-2 \ \ 4 \ \ -\mspace{-2mu}4 \ \ 2]^\rmT.$ 
    For all $k \in [0, 300],$ $c)$ shows $x_k$ for all $x_0 \in \{-4, -3, \ldots, 4\} \times \{4\} \times \{-4, -3, \ldots, 4\} \times \{2\}.$  
    $d)$ is a magnified version of $c).$
    For all $k \in [200, 300],$ the gray lines follow the trajectory from each initial state. 
    Note that each state trajectory is bounded and does not converge.
    }
    \label{fig:ex_lure_ub_3_SES}
\end{figure}
\end{exam}

\section{Conclusions and Future Work}\label{sec:conclusions}

This paper considered discrete-time Lur'e models whose response is self-excited in the sense that it is 1) bounded for all initial conditions, and 2) nonconvergent for almost all initial conditions.
These models involve asymptotically stable linear dynamics with a washout filter connected in feedback with a piecewise-C$^1$ affinely constrained nonlinearity.
Sufficient conditions involving the growth rate of the nonlinearity were given under which the system is self-excited.
Future work will focus on the following objectives:
1) motivated by Example \ref{ex_lure_ub_3}, determine whether or not LMI feasibility is a necessary condition for \eqref{xLure}, \eqref{yLure} to be self-excited;
and 2) derive analogous results for continuous-time Lur'e models.

\section*{Acknowledgments}

The first author was supported by NSF grant CMMI 1634709.

\begin{appendix}

{\it Proof of Proposition \ref{propequilibrium}.}

To prove $i)$, note that, since $I-A$ is nonsingular, it follows that \eqref{xeeqn} and \eqref{xeG0} are equivalent.

To prove $ii)a)$, note that $i)$ implies 
$C x =  C (I - A)^{-1} B (\phi(C x) + v) =  G(1) (\phi(C x) + v).$

To prove necessity in $ii)b)$, note that \eqref{xeG0} implies $x = 0.$
To prove sufficiency in $ii)b)$, note that \eqref{xeG0} implies $B(\phi(C x) + v)=0.$
Since $B$ is nonzero, it follows that $\phi(C x) = -v.$

To prove $ii)c)$, note that, since $G(1) = 0,$ it follows that $ii)a)$ implies $C x = G(1) (\phi(Cx) + v) = 0.$
Furthermore, since $I - A$ is nonsingular, \eqref{xeG0} implies that $x = (I - A)^{-1} B (\phi(0) + v)$ is the unique equilibrium of \eqref{xLure}, \eqref{yLure}.

To prove $ii)d)$, note that, since $Cx = 0,$ it follows from $ii)a)$ that $G(1) (\phi(0) + v) = 0,$ which implies that either $G(1) = 0$ or $v = -\phi(0).$

To prove $ii)e)$, note that, since $\phi(Cx) = 0,$ \eqref{x_rme_def} and \eqref{xeG0} imply $x =x_\rme.$

To prove $iii)$, note that \eqref{CxeisG1v} implies $iii)b)$ $\Longleftrightarrow$ $iii)c).$
Next, we show that $iii)a)$ $\Longrightarrow$ $iii)b)$ and $iii)b)$ $\Longrightarrow$ $iii)a).$
To prove $iii)a)$ $\Longrightarrow$ $iii)b),$ note that \eqref{xeG0} implies
$x_\rme = (I - A)^{-1} B (\phi(C x_\rme) + v) = (I-A)^{-1} B v,$ which implies $\phi(C x_\rme) = 0.$
To prove $iii)b)$ $\Longrightarrow$ $iii)a),$ note that $x_\rme  = (I-A)^{-1} B v = (I-A)^{-1} B (\phi(C x_\rme) + v).$
Hence, $i)$ implies $x_\rme$ is an equilibrium.

$iv)$ follows from $iii)$ in the case $G(1) \ne 0.$

To prove $v)$, we show
$v)c)$ $\Longrightarrow$ $v)b)$ $\Longrightarrow$ $v)a)$ $\Longrightarrow$ $v)c)$.
$v)c)$ $\Longrightarrow$ $v)b)$ is immediate.
Next, since $G(1) = 0,$ $iv)$ implies $C x_\rme = G(1) v =  0.$
Hence, $iii)$ with $C x_\rme = 0$ implies $v)b)$ $\Longrightarrow$ $v)a)$.
Finally, since $G(1) = 0,$ $ii)$ $c)$ implies that $x = (I-A)^{-1} B (\phi(0) + v)$ is the unique equilibrium of \eqref{xLure}, \eqref{yLure}.
In the case $\phi(0) = 0,$ $x = (I-A)^{-1} B v = x_\rme$ is the unique equilibrium of \eqref{xLure}, \eqref{yLure}, and thus $v)a)$ $\Longrightarrow$  $v)c)$.
\hfill $\square$

\end{appendix}

\bibliographystyle{IEEEtran}
\bibliography{IEEEabrv,bib_paper.bib}

\section*{Author Biographies}

\noindent {\bf Juan Paredes} received the B.Sc. degree in mechatronics engineering from the Pontifical Catholic University of Peru and a M.Eng. degree in aerospace engineering from the University of Michigan in Ann Arbor, MI. He is currently a PhD candidate in the Aerospace Engineering Department at the University of Michigan.  His interests are in autonomous flight control of unmanned aerial vehicles and stabilization of combustion instabilities.

 \medskip

\noindent {\bf Omran Kouba} received the Sc.B. degree in Pure Mathematics from the  University of Paris XI and the Ph.D. degree in Functional Analysis from Pierre and Marie Curie University in Paris, France. Currently, he is a professor in the Department of Mathematics in the Higher Institute of Applied Sciences and Technology, Damascus (Syria).  His interests are in real and complex analysis, inequalities, and problem solving.

 \medskip

\noindent {\bf Dennis S. Bernstein} is a faculty member in the Aerospace Engineering Department at the University of Michigan.  His interests are in identification, estimation, and control.

\end{document}